\documentclass[10pt,letterpaper]{article}
\usepackage[top=0.85in,left=2.75in,footskip=0.75in]{geometry}

\pdfoutput=1
 
\usepackage{changepage}

\usepackage[utf8]{inputenc}

\usepackage{textcomp,marvosym}

\usepackage{fixltx2e}

\usepackage{amsmath,amssymb}

\usepackage{cite}

\usepackage{nameref,hyperref}

\usepackage[right]{lineno}

\usepackage{microtype}
\DisableLigatures[f]{encoding = *, family = * }

\usepackage{rotating}

\usepackage{setspace}

\raggedright
\usepackage[aboveskip=1pt,labelfont=bf,labelsep=period,justification=raggedright,singlelinecheck=off]{caption}

\makeatletter
\renewcommand{\@biblabel}[1]{\quad#1.}
\makeatother

\date{}

\usepackage{lastpage,fancyhdr,graphicx}
\usepackage{epstopdf}
\fancyheadoffset[L]{2.25in}
\fancyfootoffset[L]{2.25in}

\usepackage{color}

\usepackage{xspace}
\newcommand{\eg}{e.g.,\xspace} %
\newcommand{\ie}{i.e.,\xspace} %
\newcommand{\fig}{Fig\xspace} %
\newcommand{\figs}{Figs\xspace} %
\newcommand{\Eqs}{Eqs\@\xspace} %

\begin{document}
\vspace*{0.2in}

\vspace{-.2in}
\begin{flushleft}
{\Large
\textbf\newline{Interference of Neural Waves in Distributed \\ \vspace{.09in} Inhibition-stabilized Networks}
}
\newline
\\
Sergey Savel'ev\textsuperscript{1},
Sergei Gepshtein\textsuperscript{2,*}
\\
\vspace{.05in}
{\bf{1}} Department of Physics, Loughborough University \\ Leicestershire, LE11 3TU, United Kingdom
\\
\vspace{.05in}
{\bf{2}} Center for Neurobiology of Vision, Salk Institute for Biological Studies \\ 10010 North Torrey Pines Road, La Jolla, CA 92037, USA

%
%




\vspace{.1in}
* Corresponding author e-mail:  {\it s.saveliev@lboro.ac.uk} $\medskip$ or {\it sergei@salk.edu} 

\vspace{.15in}
\today

\end{flushleft}

\vspace{.1in}
\noindent {\bf Abstract.} To gain insight into the neural events responsible for visual perception of static and dynamic optical patterns,
we study how neural activation spreads in arrays of inhibition-stabilized neural networks with nearest-neighbor coupling.
The activation generated in such networks by local stimuli propagates between locations, forming spatiotemporal waves that affect the dynamics of activation generated by stimuli separated spatially and temporally, and by stimuli with complex spatiotemporal structure.
These interactions form characteristic interference patterns that make the network intrinsically selective for certain stimuli, such as modulations of luminance at specific spatial and temporal frequencies and specific velocities of visual motion.
Due to the inherent nonlinearity of the network, its intrinsic tuning depends on stimulus intensity and contrast.
The interference patterns have multiple features of  ``lateral'' interactions between stimuli, well known in physiological and behavioral studies of visual systems.
The diverse phenomena have been previously attributed to distinct neural circuits.
Our results demonstrate how the canonical circuit can perform the diverse operations in a manner predicted by neural-wave interference.

\vspace{.2in}
\noindent {\bf  Author Summary.}  We developed a framework for analysis of biological neural networks in terms of neural wave interference. Propagation of activity in neural tissue has been previously studied with regards to waves of neural activation. We argue that such waves generated at one location should interfere with the waves generated at other locations and thus form patterns with predictable properties. Such interactions between effects of sensory stimuli are commonly found in behavioral and physiological studies of visual systems, but the interactions have not been studied in terms of neural wave interference. Using a canonical model of the inhibitory-excitatory neural circuit, we investigate interference of neural waves in one-dimensional chains and two-dimensional arrays of such circuits with nearest-neighbor coupling.  We define conditions of stability in such systems with respect to corrugation perturbations and derive the control parameters that determine how such systems respond to static, short-lived, and moving stimuli. We demonstrate that the interference patterns generated in such networks endow the system with many properties of biological vision, including selectivity for spatial and temporal frequencies of intensity modulation, selectivity for velocity, “lateral” interactions between spatially and temporally separate stimuli, and predictable delays in response to static and moving stimuli.

\newpage
\tableofcontents
\listoffigures
\newpage 

\doublespacing

\section*{Introduction}
\addcontentsline{toc}{section}{Introduction}
Waves propagating through a medium from different sources may form patterns. When the wave equations are linear, the waves interfere constructively or destructively, yielding local nodes and antinodes that may retain their spatial positions (standing waves), or evolve in space and time (running waves), producing dynamic patterns. Such effects were originally studied for acoustic and light waves \cite{Landau1987,landau2013classical,everest2001master,rossing1999light} followed by observations of interference for quantum particles \cite{Landau1965quantum,ficek2005quantum}.

More recently, propagation of activity in neural tissue was studied in terms of waves of neural activation, \eg \cite{ermentrout1998neural,bressloff2011spatiotemporal}.
The neural waves generated at one location in the neural tissue are expected to interfere with the waves generated at other locations and thus form patterns with predictable properties.
Such ``lateral'' interactions between effects of sensory stimuli are commonly found in psychophysical and physiological studies of biological sensing.
For example, physiological and behavioral studies of visual systems found that effects of spatially separated optical patterns interact with one another over distance \cite{polat1993lateral,polat1994architecture,kovacs1994perceptual,adini1997excitatory}.
These interactions have not been studied in terms of neural wave interference \cite{SavelievGepshtein2014entropy}.

The lateral interactions are often explained in terms of  two components of biological sensory systems.   
The first component is the specialized local neural circuits. 
The circuits are specialized in the sense they are selective for certain stimuli, which is why they are commonly modeled using the formalism of \emph{linear filter}, \eg \cite{ratliff1965mach,CampbellRobson1968,marr1982vision,pollen1983visual,daugman1985uncertainty,devalois1988spatial}. 
The second component is the long-range axonal connections between the specialized circuits discovered by anatomical and physiological methods, \eg \cite{gilbert1996spatial,bringuier1999horizontal,kapadia2000spatial,stettler2002lateral,kozyrev2014voltage}. 

In models of neural circuits as linear filters, the response to a sensory stimulus is computed by convolving the stimulus with a kernel whose parameters are estimated in psychophysical  \cite{manahilov1995spatiotemporal,manahilov1998triphasic, manahilov1999energy} or physiological \cite{jones1987evaluation,devalois1988spatial} studies, or are selected because they are suitable for generic visual tasks, \eg \cite{marr1982vision}.  
Such linear models are commonly enriched by incorporating serial  nonlinear processing stages to account for nonlinear behavior of visual systems \eg \cite{shapley1985spatial,heeger1992normalization,simoncelli1998model,rust2006MT,vintch2015convolutional}.

Here we entertain a different approach. 
We describe the network using a system of linear spatiotemporal differential equations derived from a model of the canonical neural circuit \cite{wilson_cowan1973}. 
Solutions of the system of differential equations describe the network activity evoked by stimulation. 
We analyze the linear solutions of these equations using the method of Green's function \cite{riley2006mathematical} (\emph{cf.}~\cite{poggio1990regularization,stevens1994form}). 
A Green's function represents the distribution of activity in the network generated by a small (``point'') stimulus  of unit intensity.  
The neural activity evoked by complex stimuli can be modeled, in linear approximation, by convolving the Green's function of the network with the stimulus. 
We demonstrate how  this approach allows for highly-flexible context-dependent tuning of biological sensory systems, supplanting the aforementioned phenomenological models. 

Using the model of canonical neural circuit in the inhibition-stabilized regime \cite{tsodyks1997paradoxical,ozeki2009inhibitory,ahmadian2013analysis}, we studied the interference patterns generated in arrays of neural circuits.  
The interference patterns have a number of properties that resemble properties of biological vision, but which have been often attributed to distinct specialized neural mechanisms.  
The distinct mechanisms include selectivity for spatiotemporal frequency and velocity of stimuli, ``lateral'' interactions between effects of spatially separate stimuli  \cite{polat1993lateral,field1993contour,kovacs1994perceptual,adini1997excitatory}, 
and delays in perception of static and dynamic stimuli \cite{nijhawan2002neural,krekelberg2001neuronal,hubbard2005representational}.  
We propose that the diverse phenomena can arise from the same canonical circuit, and that they depend on a small number of control parameters that allow neural systems to flexibly maintain their tuning to useful properties of the environment. 

In particular, our analysis reveals which features of  network connectivity determine its tuning to optical stimuli and thus define its \emph{intrinsic} filtering characteristics.   
The notion of intrinsic tuning offers an alternative to the approaches that require phenomenological models of neural filters. 
We show further that at low stimulus contrasts the spatial filters derived this way behave similar to linear filters, in agreement with an assumption common in models of visual mechanisms operating near the threshold of visibility
\cite{wilson1979four,shapley1985spatial,van1992information}. 
Yet the filtering properties of these networks change at higher stimulus contrasts.  
Thus, the network's intrinsic spatial-frequency tuning shifts in a manner predictable from network parameters, again
offering an alternative to the phenomenological models of system nonlinearities (\emph{cf.}~\cite{ahmadian2013analysis,rubin2015stabilized}).   
We demonstrate how the  framework of neural-wave interference applies in one-dimensional chains and two-dimensional arrays of neural units.  
We show how this framework helps to understand the perception of two-dimensional stimulus configurations in terms of  spreading neural activation and interference of the neural waves converging from multiple locations.

\section*{Neural chain with nearest-neighbor coupling}
\addcontentsline{toc}{section}{Neural chain with nearest-neighbor coupling}

\subsection*{Model of inhibition-stabilized neural chain}
\addcontentsline{toc}{subsection}{Model of inhibition-stabilized neural chain}

Neural networks underlying sensory processes have been modeled at different levels of abstraction, focusing on  local circuitry or on  interactions between the local circuits \cite{bressloff2011spatiotemporal,Wilson1999book}. 
Yet the consequences of interference tend to depend more significantly on network geometry and topology (\ie on cell connectivity, network dimensionality, on whether the network is fractal, etc) than on the finer detail of local circuitry: the ``node'' of the network. 
Here we investigate basic principles of neural-wave interference using a repetitive canonical neural motif: an inhibition-excitation node, forming one-dimensional chains or two-dimensional arrays. 

\begin{figure}[h!]
	\begin{center}
		\includegraphics[width=\textwidth]{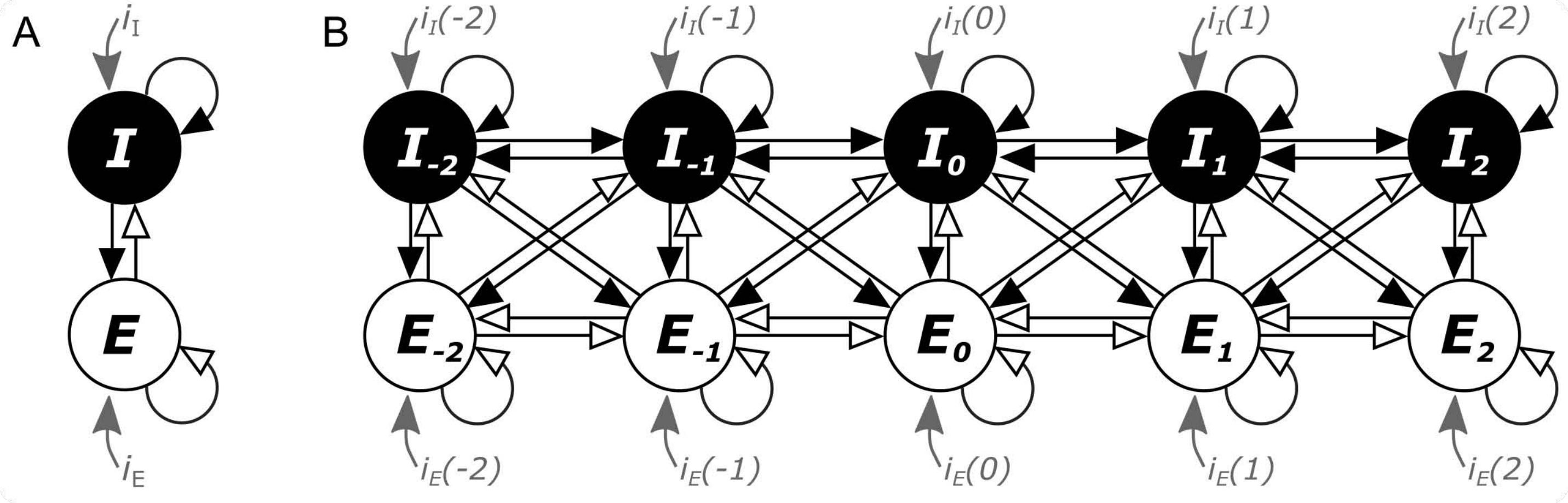} 
	\end{center}
	\caption[Canonical circuit and neural chain]{\small Canonical circuit and neural chain. 
		(A)~The elementary Wilson-Cowan circuit contains two reciprocally connected cells: excitatory ($E$) and inhibitory ($I$), both with recurrent feedback. 
		The arrows represent excitatory and inhibitory connections, respectively as the blank and filled heads. 
		%
		(B)~Several canonical circuits are arranged as ``nodes" in a chain with nearest-neighbor coupling. 
		Integer indices $l$ 
		indicate  node locations in the chain, $l \in \{-2, ..., 2\}$. Currents $i_{E}(l)$ and $i_I(l)$ are the inputs into each node generated by the stimulus. 
	}
	\label{fig:ISN}
\end{figure}

We use the Wilson-Cowan model of the canonical circuit \cite{wilson_cowan1973} because even a single-node version of it (\fig~\ref{fig:ISN}A) has proven useful for understanding behavior of small cortical circuits \cite{ozeki2009inhibitory}, and because its rich dynamical properties \cite{tsodyks1997paradoxical} have been helpful in studies of larger-scale neural phenomena \cite{bragin1995gamma,jadi2014regulating}.


To begin, consider a nearest-neighbor chain of Wilson-Cowan nodes \cite{wilson_cowan1973,Wilson1999book} illustrated in \fig~\ref{fig:ISN}B.
The dynamics of this network is described by a system of equations
\begin{eqnarray}
\tau_E\frac{dr_E(l)}{dt}&=&-r_E(l) + g({\cal W}_E)\nonumber \\
\frac{dr_I(l)}{dt}&=&-r_I(l) +g({\cal W}_I), \nonumber \\
\label{main-chain}
\end{eqnarray}
with the weights of connections between the cells represented by the excitatory ($E$) and inhibitory ($I$) coefficients:
\begin{eqnarray}
{\cal W}_E = w_{EE}r_E(l)+{\tilde w}_{EE}r_E(l+1)+{\tilde w}_{EE}r_E(l-1)\nonumber \\
-w_{EI}r_I(l)-{\tilde w}_{EI}r_I(l+1)-{\tilde w}_{EI}r_I(l-1)+i_E(l),
\label{main-chain_weights_E}
\end{eqnarray}
\begin{eqnarray}
{\cal W}_I = w_{IE}r_E(l)+{\tilde w}_{IE}r_E(l+1)+{\tilde w}_{IE}r_E(l-1)\nonumber \\
-w_{II}r_I(l)-{\tilde w}_{II}r_I(l+1)-{\tilde w}_{II}r_I(l-1)+i_I(l),
\label{main-chain_weights_E}
\end{eqnarray}
where 
\begin{itemize}
	\item $l$ is the number (or, equivalently, the discrete spatial coordinate) of an excitatory-inhibitory node in the chain, 
	\item $g(\cdot)$ is a sigmoid function characterizing the firing rate of an individual cell as a function of its input intensity,
	\item  $r_E(l)$ and $r_I(l)$ are the firing rates of, respectively, the excitatory and inhibitory cells at node $l$,
	\item $\tau_E$ is the characteristic time of excitatory cells,   
	measured in the units of characteristic time of relaxation of the inhibitory cell.
\end{itemize}
The coefficients $w_{EE}$, $w_{EI}$, $w_{IE}$ and $w_{II}$ describe the interactions of the excitatory and inhibitory cells within every node (as in \cite{ozeki2009inhibitory}), and ${\tilde w}_{EE}$, ${\tilde w}_{EI}$, ${\tilde w}_{IE}$, ${\tilde w}_{II}$ represent the strength of coupling between the nearest nodes.  The inputs $i_E(l)$ and $i_I(l)$ of the excitatory and inhibitory cells are provided by the optical stimulation, while the input ratio 
$\alpha = i_E(l)/(i_E(l)+i_I(l))$ 
is the same for all the nodes (and thus it could be modeled as an additional parameter). 

We first investigate the linear regime of this system, and then consider a nonlinear case. 
In the simpler linear case, we approximate the sigmoid function $g$ as $g(x)\approx x$. 
In other words, we consider the following equations:
\begin{eqnarray}
\tau_E\frac{dr_E(l)}{dt}&=&-r_E(l) + {\cal W}_E\nonumber \\
\frac{dr_I(l)}{dt}&=&-r_I(l) +{\cal W}_I. \nonumber \\
\label{main-lin-chain}
\end{eqnarray}

\subsection*{Stability of neural chain}
\addcontentsline{toc}{subsection}{Stability of neural chain}
\label{sec:stability}

\noindent We begin by analyzing the stability of the linearized equations, since a similar analysis of two coupled equations for a single node has been helpful for understanding the dynamical regimes observed in local circuits: pairs of coupled excitatory and inhibitory cells \cite{tsodyks1997paradoxical}. 

Consider a perturbation around the solution $r_E(t,l)=r_I(t,l)=0$ at zero inputs $i_E=i_I=0$. In contrast to the single-node configuration \cite{tsodyks1997paradoxical}, we must take into account the spatial dependence of the perturbations, namely
\begin{eqnarray}
r_E=\Delta_E e^{\lambda t+ik l} \nonumber \\
r_I= \Delta_I e^{\lambda t +ik l}
\label{chain-perturb}
\end{eqnarray}
with constant amplitudes $\Delta_E$ and $\Delta_I$, spatial wave number $k$, and temporal rate $\lambda$. 
By substituting (\ref{chain-perturb}) in~(\ref{main-chain}) we observe:
\begin{eqnarray}
(w_{EE}+2{\tilde w}_{EE}\cos k-1-\tau_E\lambda)\Delta_E-(w_{EI}+2{\tilde w}_{EI}\cos k)\Delta_I=0, \nonumber \\
(w_{IE}+2{\tilde w}_{IE}\cos k)\Delta_E-(w_{II}+2{\tilde w}_{II}\cos k +1+\lambda)\Delta_I=0.
\label{chain-perturb1}
\end{eqnarray}
This algebraic set has a solution if the temporal rate of $r_E$ is
\begin{equation} 
\lambda_{\pm}=\frac{{\bar W}_{EE}-1-\tau_{E}{\bar W}_{II}-\tau_E\pm\sqrt{\cal A}}{2\tau_E} 
\label{lampm},
\end{equation}
where
\begin{equation}
{\cal A} = ({\bar W}_{EE}-1-\tau_{E}{\bar W}_{II}-\tau_E)^2-4\tau_E[({\bar W}_{II}+1)(1-{\bar W}_{EE})+{\bar W}_{EI}{\bar W}_{IE}],
\label{term_A}
\end{equation}
with
\[
{\bar W}_{s}=w_{s}+2{\tilde w}_{s}\cos k,
\]
and index $s$ runs through values $EE,EI,IE,II$. Note that results of \cite{tsodyks1997paradoxical} follow when all values of ${\tilde w}_{s}$ take zero.

Stability of the neural chain requires that all spatial-temporal activations decay in time,  which happens if  ${\rm Re}\lambda_{\pm}(k)<0$ for all $k$, resulting in conditions
\begin{eqnarray}
w_{EE}-1-\tau_{E}w_{II}-\tau_E+2{\cal R}\cos k<0\ \ {\rm and} \nonumber \\
(w_{II}+1)(1-w_{EE})+w_{EI}w_{IE}-{\cal K}\cos^2 k-2{\cal KT}\cos k >0,
\label{chain-stab-cond}
\end{eqnarray}
where
\begin{equation}
{\cal K}=4({\tilde w}_{II}{\tilde w}_{EE} -{\tilde w}_{EI}{\tilde w}_{IE}), \ \ \ {\cal R}={\tilde w}_{EE}-\tau_{E}{\tilde w}_{II},
\label{kk}
\end{equation}
and
\begin{equation}
{\cal T}=({\tilde w}_{EE}(w_{II}+1)+{\tilde w}_{II}(w_{EE}-1)-{\tilde w}_{EI}w_{IE}-{\tilde w}_{IE}w_{EI})/{\cal K}.
\label{tt}
\end{equation}
Since $|\cos k|<1$, the first stability equation in (\ref{chain-stab-cond}) can be rewritten as
\begin{equation}
{\cal Q}=w_{EE}-1-\tau_{E}w_{II}-\tau_E+2|{\tilde w}_{EE}-\tau_{E}{\tilde w}_{II}|<0,
\label{condition1}
\end{equation}
where the slowest decaying perturbation has the wave number $k=0$  
if ${\cal R}={\tilde w}_{EE}-\tau_{E}{\tilde w}_{II}>0$ or $k\approx \pi$ if ${\cal R}={\tilde w}_{EE}-\tau_{E}{\tilde w}_{II}<0$ for ${\cal A}<0$.

When ${\cal A}<0$ and the chain is stimulated by a short pulse of current, the activation can persist for a long while with the spatial wavelengths of either $k \approx 0$ (all nodes oscillate in-phase) or $k\approx \pi$ (neighboring nodes oscillate out-of-phase) when, respectively,  ${\cal R}>0$ or ${\cal R}<0$; much longer than for the spatial wavelengths different from $0$ and $\pi$.
This result is consistent with the numerical simulations described in Section~{\it Response to a short-lived stimulus} (p.~\pageref{sec:short_lived_simulations}), illustrated in \figs~\ref{fig:spatiotemporal:new}-\ref{fig:spatiotemporal:new2}. 

Analyzing the second stability condition in (\ref{chain-stab-cond}), we obtain parameter regions where the set of equations (\ref{main-lin-chain}) is stable. Namely, in addition to the constraint (\ref{condition1}), the following conditions should be satisfied:
\begin{eqnarray}
(w_{II}+1)(1-w_{EE})+w_{EI}w_{IE}-{\cal K}+2{\cal KT}>0\ \ {\rm if}\ \ {\cal K}>0,\ {\cal T}<0,
\nonumber \\
(w_{II}+1)(1-w_{EE})+w_{EI}w_{IE}-{\cal K}-2{\cal KT}>0\ \ {\rm if}\ \ {\cal K}>0,\ {\cal T}>0, \nonumber \\
(w_{II}+1)(1-w_{EE})+w_{EI}w_{IE}-{\cal K}+2{\cal KT}>0\ \ {\rm if}\ \ {\cal K}<0,\ {\cal T}>1,
\nonumber \\
(w_{II}+1)(1-w_{EE})+w_{EI}w_{IE}-{\cal K} -2{\cal KT}>0\ \ {\rm if}\ \ {\cal K}<0,\ {\cal T}<-1,
\nonumber \\
(w_{II}+1)(1-w_{EE})+w_{EI}w_{IE}+{\cal K}{\cal T}^2>0\ \ {\rm if}\ \ {\cal K}<0,\ -1<{\cal T}<1.
\label{chain-second-stability}
\end{eqnarray}
These equations define the regime of stabilization by inhibition, including the terms responsible for stabilization of individual nodes (the first additive terms in every row) and the terms responsible for stabilization of the chain. 
As mentioned, under null interactions between the nodes (i.e., under ${\cal K}=0,\ {\cal KT}=0$) we reproduce the results derived in \cite{tsodyks1997paradoxical}. 

In the chain, the instability  can occur over time because of an exponential increase of spatial amplitudes 
$ \Delta_{E}\exp(-\lambda(k) t) $
and 
$ \Delta_I\exp(-\lambda(k) t)$ 
of the respective waves $r_E(t,l)$ and $r_I(t,l)$ with different wave numbers $k$ (i.e., with the spatial periods of $2\pi/k$). 
This so-called \emph{corrugation instability} is well known in acoustics and hydrodynamics \cite{Landau1987}. 
Here it arises because of the mutual excitation of nodes in the chain. 

If either (\ref{condition1}) or (\ref{chain-second-stability}) is not satisfied, the solution $r_E(t,l)=r_I(t,l)=0$ at $i_E(l)=i_I(l)=0$ becomes unstable. 
In this case, the system can be attracted to another fixed point, or it can develop periodic, quasi-periodic, or chaotic oscillations. 
This behavior is different from the behavior of the single-node model \cite{tsodyks1997paradoxical} because in our case the oscillations occur in both space and time.
The oscillations can be chaotic by virtue of mixing the waves with different wave numbers $k$, producing oscillations at different time scales $\lambda_{\pm}(k)$ and different temporal frequencies. 
In what follows, we focus on the interference of neural waves in the regime of stability~(\ref{condition1})--(\ref{chain-second-stability}).

\subsection*{Parameters responsible for behavior of neural chain}
\addcontentsline{toc}{subsection}{Parameters responsible for behavior of neural chain}

\noindent To summarize, the preceding analysis of \Eqs~\ref{main-lin-chain} helped to reveal the parameters that underlie qualitatively different classes of network behavior. 
The control parameters are:
\begin{eqnarray*}
	{\cal K}=4({\tilde w}_{II}{\tilde w}_{EE} -{\tilde w}_{EI}{\tilde w}_{IE}), {\cal R}={\tilde w}_{EE}-\tau_{E}{\tilde w}_{II},
	\label{kk}
\end{eqnarray*}
\begin{equation*}
{\cal T}=({\tilde w}_{EE}(w_{II}+1)+{\tilde w}_{II}(w_{EE}-1)-{\tilde w}_{EI}w_{IE}-{\tilde w}_{IE}w_{EI})/{\cal K}.
\label{tt}
\end{equation*}
\begin{equation*}
{\cal Q}=w_{EE}-1-\tau_{E}w_{II}-\tau_E+2|{\tilde w}_{EE}-\tau_{E}{\tilde w}_{II}|<0,
\label{condition1_1}
\end{equation*}
\begin{equation*}
{\cal M}=(w_{II}+1)(1-w_{EE})+w_{EI}w_{IE}+{\cal K T}^2.
\end{equation*}

Although this modeling framework employs a large overall number of parameters, all stationary solutions (the spatial distribution of network activity) are controlled by two quantities, ${\cal T}$ and ${\cal M/K}$, while network dynamics is additionally controlled by ${\cal Q}$ and ${\cal R}$. 

Remarkably, when we choose different sets of parameters that correspond to the same magnitudes of ${\cal T}$ and ${\cal M/K}$ but different magnitudes of ${\cal R}$, the network produces nearly the same responses for lasting stimuli, and very different responses for short-lived stimuli. We illustrate this behavior in the numerical simulations described in Section~{\it Spatiotemporal interference} (p.~\pageref{sec:spatiotemp}).

\section*{Static neural waves generated by point stimulus}
\addcontentsline{toc}{section}{Static neural waves generated by point stimulus}
\label{sec:static_point}

\noindent 
Consider a ``point'' visual stimulus that is small enough to directly affects only one node of the network. 
The activation produced by such a point stimulus  propagates through the chain by means of the nearest-neighbor coupling.

The steady-state response $r_E(l)$ of the chain to a point stimulus can be modeled as a static excitation that generates currents $i_E$ and $i_I$:
\[
i_E(l\ne 0)=i_I(l\ne 0)=0,\ i_E(0)=\alpha j_0, \ i_I(0)=(1-\alpha)j_0,
\]
where the current  $j(l=0)=j_0$ is applied to the zeroth node alone, \ie $j(l\ne 0)=0$.
To simplify analysis, we associate this solution with the spatial Green's function for $j_0=1$.  

\begin{figure}[t!]
	\begin{minipage}{.63\textwidth}
		\centering
		\includegraphics[width=\textwidth]{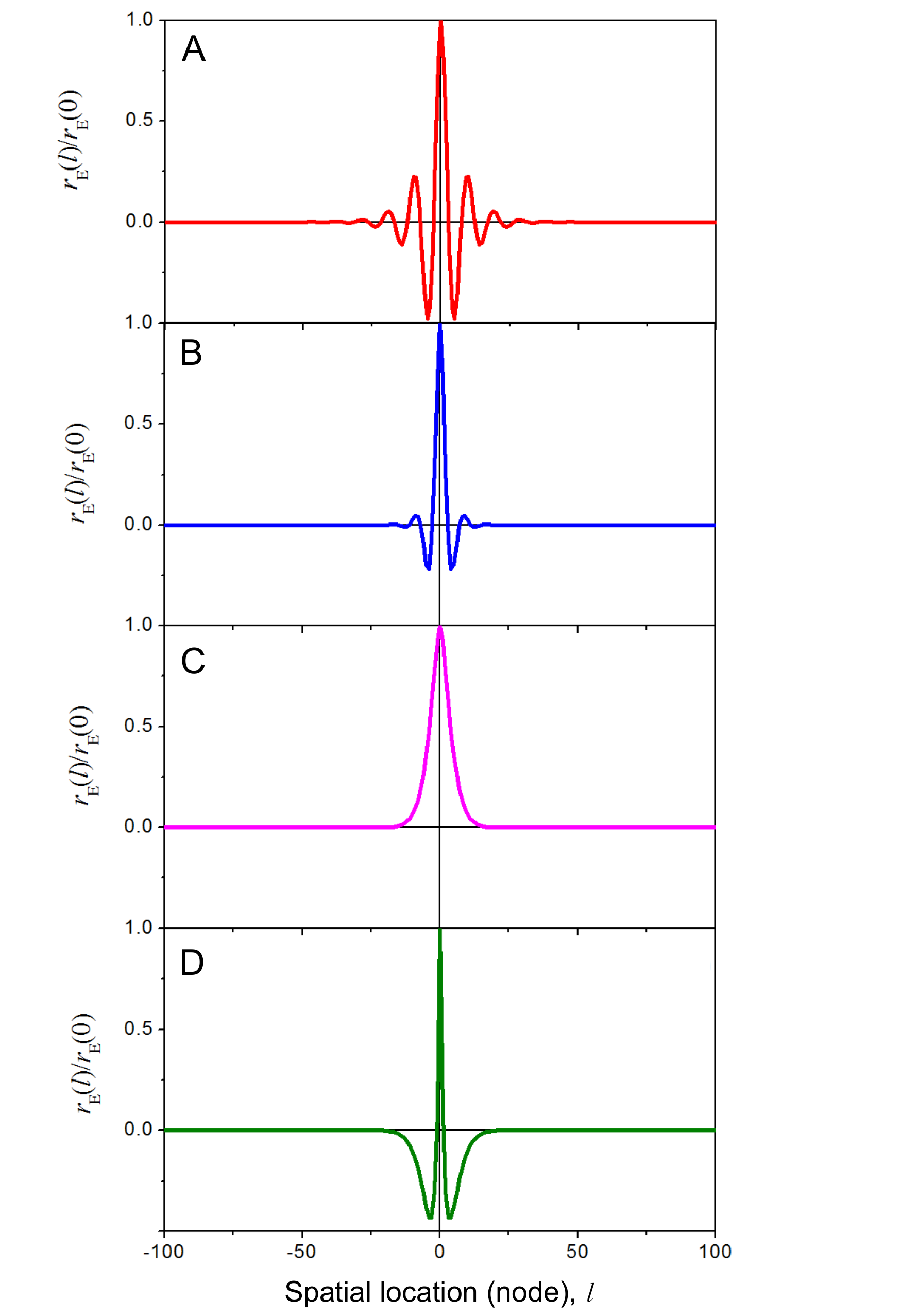} 
	\end{minipage}
	\hfill
	\begin{minipage}[c]{.37\textwidth}
		\caption[Green's functions]{\small	
			Stationary response 
			to a static ``point'' stimulus calculated using (\ref{main-lin-chain}) with $i_E(0)=\alpha j_0$ and  $i_I=(1-\alpha)j_0$, where $j_0=1$, $\tau_E=4$, $w_{EE}=2$, $w_{IE}=1.5$, ${\tilde w}_{EE}={\tilde w}_{IE}={\tilde w}_{EI}=1$. Chain length was 200 nodes. The simulation time required to reach the steady state was 10,000 with the time step $dt=0.0001$.  
			Other parameters were: 
			(\textbf{A})~${\tilde w}_{II}=0.7$, ${\cal T}=-0.8$ and ${\cal M}=0.01$, $\alpha=0.8$; 
			(\textbf{B})~same as~A except ${\tilde w}_{II}=0.4$ and ${\cal M}=0.1$;
			(\textbf{C})~same as~A except ${\cal T}=1.1$;
			(\textbf{D})~same as~C except $\alpha=0.461$.
		} 
		\label{fig:spatial_greens_functions} 
	\end{minipage}
\end{figure}

Given different magnitudes of the control parameters ${\cal T}$, ${\cal M/K}$, and $\alpha$, the response can take the different forms illustrated in \fig~\ref{fig:spatial_greens_functions}: spatially damped oscillations~(panels~A and~B), purely exponential spatial decay~(C), or two competing exponents decreasing away from the stimulated node and generating two negative minima near the central positive maximum~(D).

The numerical results summarized in \fig~\ref{fig:spatial_greens_functions} can be readily confirmed by analysis of the set of equations~(\ref{main-lin-chain}).  
Indeed, for every node in the chain, except for the directly activated node (\ie for all $l\ne 0$), we substitute 
\[r_E=\Delta_E e^{ik |l|}, r_I=\Delta_I e^{ik |l|}\]
into the linear equations~\ref{main-lin-chain}, with $\Delta_E$, $\Delta_I$ and $k$ being constant  in both space and time. We thus obtain a homogeneous algebraic set (\ref{chain-perturb1}) where $\lambda$ is set to zero. A non-zero solution for $\Delta_E$  and $\Delta_I$ in these equations requires the zero determinant of the set, which can be rewritten as
\begin{equation}
{\cal M}-{\cal K}\left(\cos k+{\cal T}\right)^2=0,
\label{chain-stat-det}
\end{equation}
and which in  turn allows one to estimate the parameter $k$ responsible for both the intrinsic wavelength of the chain and for the spatial decay rate.

As we saw in the previous section, a point stimulus is expected to generate a spatial neural field distributed across the nodes and decaying away from the stimulus. 
The decay occurs if $k$ is complex, namely if
$k=i\kappa +{\tilde k}$ for $l>0$ and $k=-i\kappa - {\tilde k}$ for $l<0$, while $\kappa>0$ to ensure that $r_E(l)\rightarrow 0$ for $l\rightarrow \pm \infty$.
Substituting into (\ref{chain-stat-det}), we derive
\begin{equation}
\cos {\tilde k}=\pm 1,\ \ \pm \cosh \kappa+{\cal T}=\pm\sqrt{\cal  M/K}
\label{no-oscil-stat}
\end{equation}
for ${\cal M/K}>0$, and
\begin{equation}
\cos{\tilde k}\cosh\kappa=-{\cal T},
\sinh \kappa \sin{\tilde k}= \pm\sqrt{-{\cal M/K}}
\label{oscil-stat}
\end{equation}
for ${\cal M/K}<0$.

One of the solutions of (\ref{no-oscil-stat}) with $\cos{\tilde k}=1$ has the form of an exponentially decaying function with two damping rates that correspond to two different signs of $\sqrt{{\cal M}/{\cal K}}$. 
These solutions are shown in \fig~\ref{fig:spatial_greens_functions}C,D.
The solution of (\ref{oscil-stat})  with small and negative ${\cal M}/{\cal K}$, $|{\cal M}/{\cal K}|\ll 1$, as well as $|{\cal T}|<1$, describes spatial damped oscillations (\fig~\ref{fig:spatial_greens_functions}A,B) with 
${\tilde k}\approx \arccos(-{\cal T}$)  and 
$\kappa\approx \sqrt{-{\cal M}/({\cal K}(1-{\cal T}^2))}$.  
The parameter~${\cal T}$ determines whether the system has a purely decaying spatial response (for $|{\cal T}|>1$) or it shows decaying spatial oscillations (for  $|{\cal T}|<1$) with wave number ${\tilde k}$ (e.g., \fig~\ref{fig:spatial_greens_functions}). 
The ratio ${\cal M}/{\cal K}$ defines the exponent of spatial decay (\ie the rate at which the response decreases along the chain) away from the stimulus.

This approach allows one to predict all possible responses of the system to point stimuli in the linear regime.  

\section*{Spatial interference of waves in the neural chain}
\addcontentsline{toc}{section}{Spatial interference of waves in the neural chain}

\noindent
When the stimulus contains spatially distinct parts, or when a spatially extensive stimulus has a complex profile, we associate small stimulus regions with different nodes of the neural chain (notated by index $l$), so different parts of the stimulus are defined as different inputs $j(l)$.
The resulting waves propagate through the chain, where they may interfere with one another.
In the linear approximation of the neural response (\ie assuming that stimulus contrast is low), the waves propagate through the system independent of one another. 
The resulting distributed  response $r_E(l)$ of the chain is the weighed sum of the waves generated by point sources with unit intensity ($j_0=1$):
$$
r_E(l)=\sum_{{\tilde l}}j({\tilde l})G_E(l-{\tilde l}),
$$ 
where $G_E(l)$ is one of the functions displayed in \fig~\ref{fig:spatial_greens_functions}.
On these assumptions, the response measured on every node of the chain 
amounts to a sum of neural waves elicited from different stimulus regions.  

We illustrate this behavior below for two generic situations. 
First, we consider two elementary ``point'' stimuli (\ie stimuli with no internal structure) presented at two distinct spatial locations. 
Second, we consider a spatially extended stimulus: a cosine grating weighted by an exponential spatial envelope (``Gabor patch''). 

The results suggest a new interpretative framework for the experimental studies of visual sensitivity to different stimulus configurations, including two-dimensional spatial configurations, \eg \cite{field1993contour,kovacs1994perceptual,adini1997excitatory,series2003silent} to which the response should be modeled using two-dimensional arrays of nodes (Section~{\it Spatial dynamics in a two-dimensional neural array}, p.~\pageref{sec:2d}). 

\subsection*{Interference of point stimuli}
\addcontentsline{toc}{subsection}{Interference of point stimuli}

\begin{figure}[t!]
	\begin{center}
		\includegraphics[width=.85\textwidth]{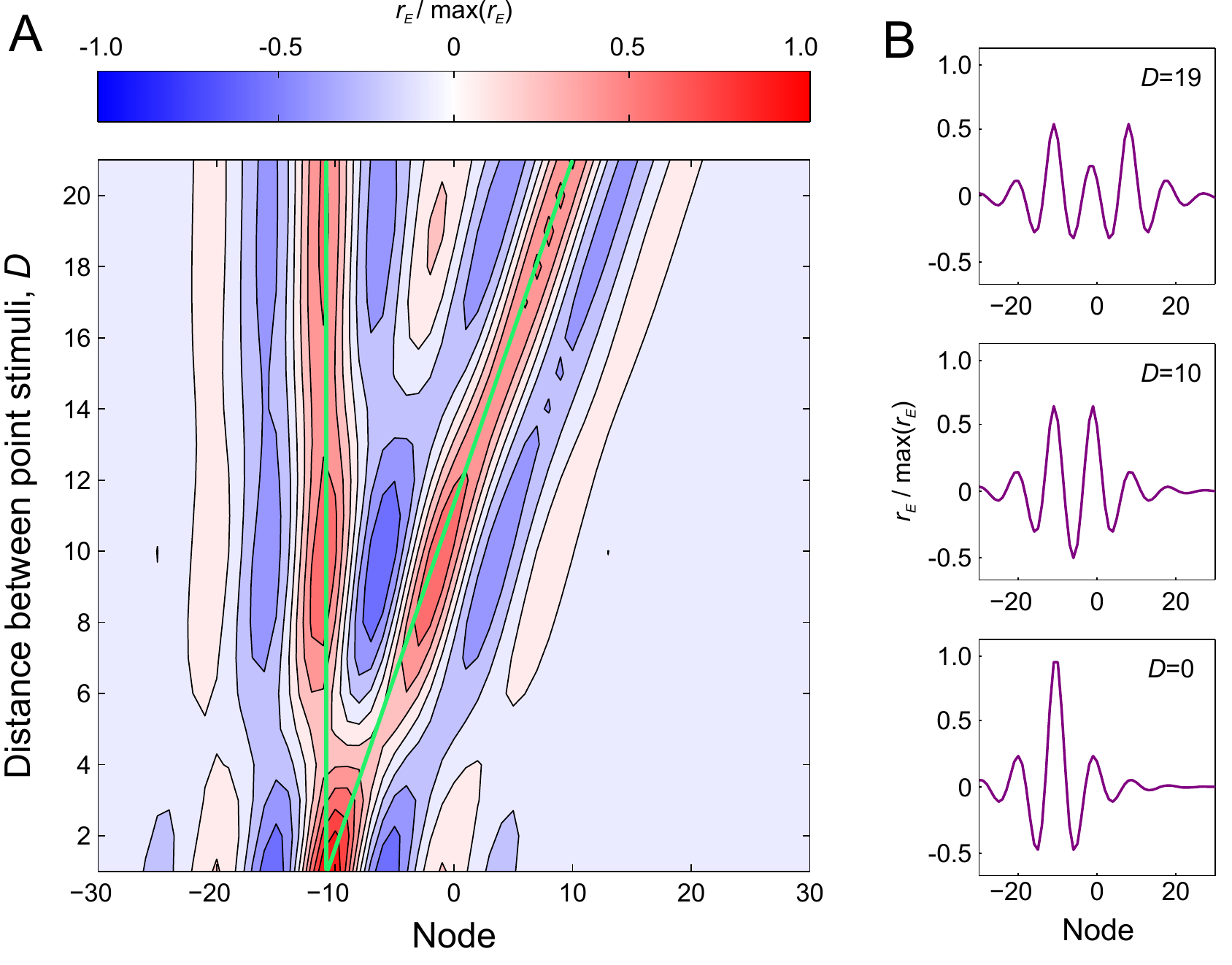}
	\end{center}
	\caption[Spatial interference of point stimuli]{\small Spatial interference of two point stimuli.
		({\bf A}) Map of responses $r_E(l)$ across location (node number) $l$ for different distances $l_1$ and $l_2$ between two static point stimuli shown in green. The ordinate represents the inter-stimulus distance $D = l_2 - l_1$.
		Because of  neural wave interference, the middle point between the stimuli is either facilitated or suppressed, depending on the inter-stimulus distance and the intrinsic spatial period of the system. 
		The map is normalized to $\max{r_E}$. 
		({\bf B}) Horizontal slices of the map in panel A: examples of the interference waveform for three inter-stimulus distances~$D$. 
	}
	\label{fig:chain_gabor}
\end{figure}

\noindent 
Suppose the chain is activated by two identical point stimuli located at positions associated with nodes $l_1$ and $l_2$. 
In other words, we consider chain response to input currents $i_E(l)=i_I(l)=0$ for any $l$ except $l_1$ and $l_2$, such that
\[
i_E(l_1)=i_E(l_2)=\alpha j_0, \ i_I(l_1)=i_I(l_2)=(1-\alpha)j_0.
\]
The response predicted by the linear set (\ref{main-lin-chain}) is summarized in \fig~\ref{fig:chain_gabor}. 
It is an interference pattern of the two ``static waves'' generated by point stimuli: 
\[r_E(l)=j_0\Bigl(G_E(l-l_1)+G_E(l-l_2)\Bigr) .\]
\fig~\ref{fig:chain_gabor}A is a map of responses for different distances $l_1-l_2$ between the stimuli (shown on the ordinate). By virtue of interference of the two neural waves, the middle position between the point stimuli  ($l=(l_1+l_2)/2$) could be facilitated  or  suppressed (\fig~\ref{fig:chain_gabor}B), depending  on whether the interference of waves generated at $l_1$ and $l_2$ is constructive or destructive.
Such predictions can be tested experimentally, for example by measuring the detection threshold of a faint point stimulus (a ``probe'') located  between the two high-contrast stimuli (``inducers'').  The contrast threshold of the probe is expected to oscillate as a function of distance between the inducers. 

\subsection*{Interference between parts of extended stimuli}
\addcontentsline{toc}{subsection}{Interference between parts of extended stimuli}

\begin{figure}[t!]
	\begin{minipage}{.4\textwidth}
		\includegraphics[width=\textwidth]{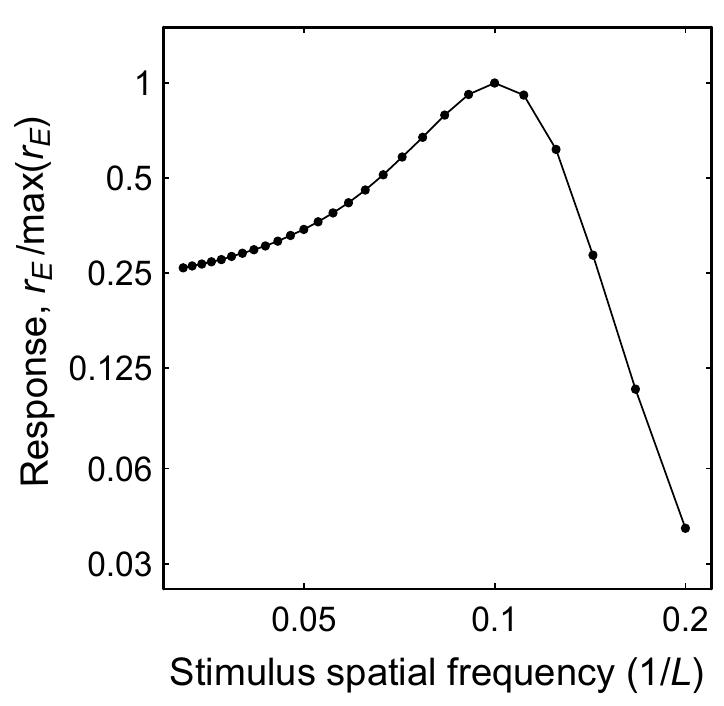}
	\end{minipage}%
	\hfill
	\begin{minipage}{.55\textwidth}
		\vspace{-5mm}
		\caption[Intrinsic tuning of the neural chain]
		{\small 
			Intrinsic tuning of the neural chain.
			Response of the chain to a distributed stimulus (luminance grating) is shown for the node $l=0$ at different spatial frequencies of the grating.  
			This response was obtained by  numerically solving equations (\ref{main-lin-chain}) with the same parameters as in \fig~\ref{fig:spatial_greens_functions}A. 
			The maximum response is attained where the stimulus spatial wavelength matches the intrinsic spatial wavelength of the chain (which is approximately eight nodes). 
			Stimulus spatial frequency is defined relative to the inter-node distance $n_0=20$.
		}
		\label{fig:intrisic:sf}
	\end{minipage}
\end{figure}

\noindent
Now consider the interference pattern created by the neural waves generated in different regions of a spatially extended stimulus,
such as the ``Gabor patch'' commonly used in physiological and psychophysical studies:
\begin{equation}
j(l)=j_0\cos(2\pi l/n_1)\exp(-l^2/n_{0}^{2}),
\label{static_gabor_patch}
\end{equation}
where $n_1$ is spatial period, $n_0$ is Gaussian width, and $j_0$ is amplitude.
As above, the network response in the linear regime can be written as convolution of the Green's function 
with the stimulus (\ref{static_gabor_patch}):
\[
r_E(l)=j_0\sum_{{\tilde l}} \cos(2\pi {\tilde l}/n_1)\exp(-{\tilde l}^2/n_{0}^{2})G_E(l-{\tilde l}).
\]
In \fig~\ref{fig:intrisic:sf}B we plot network response as a function of stimulus spatial frequency $1/n_1$.  The response is shown for the location of strongest excitation, $r_E(l=0)$. As stimulus period increases, the response first grows, reaching a maximum, and then decays. This effect can be interpreted as a constructive interference of the neural waves generated in different parts of the chain in response to the maxima and minima of stimulus luminance profile. 
The maximal response $r_E(l=0,n_1)$ is reached when the periodicity of stimulus matches the periodicity of neural waves, \ie 
when the stimulus evokes such neural oscillations in which the wavelength coincides with the spatial period of the stimulus. 
This period of oscillation may be called the ``intrinsic,'' ``natural,'' or ``characteristic'' spatial period of the network.

We conclude that, under certain conditions, the inhibition-stabilized neural chain can be characterized by the intrinsic spatial frequency (wave number) $k_s$ or by intrinsic wavelength
\begin{equation}
\lambda_s=2\pi/k_s.
\label{eq:lambda_s}
\end{equation}
In other words, the network will function as a spatial frequency filter. It will selectively  respond to some spatial frequencies in the stimulus,  consistent with the well-established notion neural cells early in the visual system are tuned to spatial frequency, \eg   \cite{hubel1968receptive,shapley1985spatial,jones1987evaluation,devalois1988spatial,priebe2006tuning}.

This view is consistent with a common description of the early stages of visual process as a bank of filters, \eg   \cite{wilson1979four,marr1982vision,adelson1985spatiotemporal}  selective for  spatial frequencies, where parameters of individual filters (or parameters of the larger system) are estimated in psychophysical, \eg  \cite{polat1993lateral,manahilov1995spatiotemporal,manahilov1998triphasic} and physiological \cite{jones1987evaluation,devalois1988spatial} studies.  
In contrast to the phenomenological models of neural filters, the present approach allows one to predict filtering properties of the network in the form of the Green's functions of the equations describing the network, and then test specific hypotheses about the neural circuitry using psychophysical methods.  

Note that, while the network is tuned to a characteristic (\textit{intrinsic}) spatial frequency, the tuning is  defined only for a certain spatial extent of the chain. In agreement with Gabor's uncertainty principle \cite{Gabor1946,Resnikoff1989,daugman1985uncertainty,GepshteinTyukinKubovy2007}, the intrinsic spatial frequency depends on the interaction of nodes in the chain, and it is not defined for a node taken alone or a system of non-interacting nodes.

\begin{figure}[t!]
	\begin{minipage}{.3\textwidth}
		\includegraphics[width = .9\columnwidth]{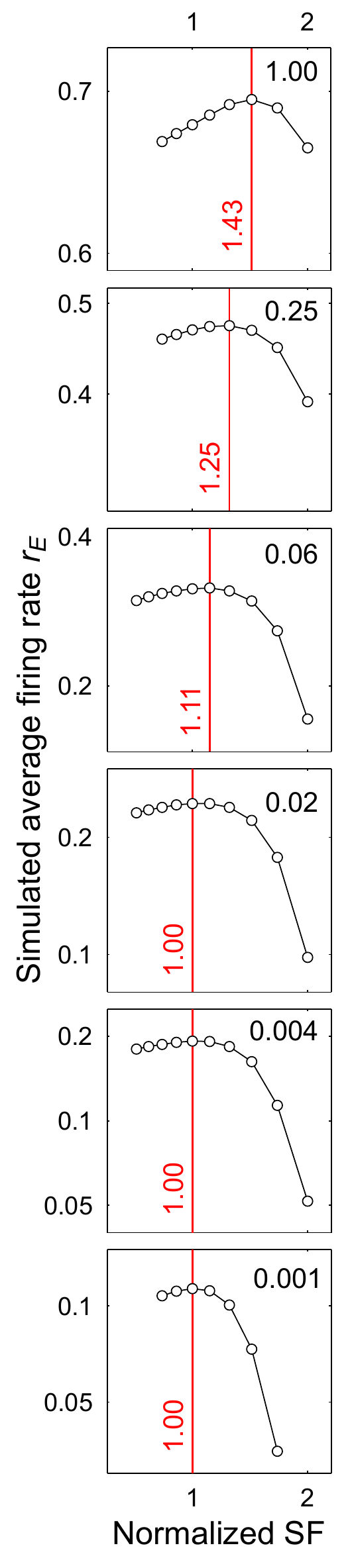}
	\end{minipage}%
	\hfill
	\begin{minipage}[c]{.6\textwidth}
		\caption[Nonlinear response]{
			Effect of stimulus strength ($j_0$) on the network response to a large static stimulus (\ref{static_gabor_patch}).		
			The response was computed using (\ref{main-chain}-\ref{main-chain_weights_E}) with the parameters as in \fig~\ref{fig:spatial_greens_functions}A. 
			Response $r_E(0)$ at stimulus center is plotted as a function of stimulus spatial frequency at six stimulus magnitudes $j_0$ 
			ranging from~0.0002 to~0.2. 
			The abscissa is the ratio of stimulus spatial frequency (``SF'') to the intrinsic spatial frequency of the network discovered using the weakest input of $j_0 = 0.0002$.  
			To interpret the inputs, we introduce a new variable $C = j_0/0.2$ called ``effective contrast''  (labeled in every panel at top right). 
			The red vertical lines mark the peak frequencies of response curves: the intrinsic frequencies of the network discovered at  different effective contrasts $C$. 
			The intrinsic frequency does not change with effective contrast while the latter is  low  (0.001 to 0.02), but then the intrinsic frequency increases with effective contrast: by 11\%, 25\%, and 43\% for respective effective contrasts of 0.06, 0.25, and 1.00 (relative to the lowest effective contrast).			
		}
		\label{fig:gabor-chain-contrast}
	\end{minipage}
\end{figure}

\subsection*{Intrinsic frequency in the nonlinear regime}
\addcontentsline{toc}{subsection}{Intrinsic frequency in the nonlinear regime}
\label{sec:intrisic:freq}

\noindent
Above we investigated behavior of neural chains in linear approximation, which we assumed to hold at low stimulus contrasts.
The linear approximation is expected to break down as stimulus contrast is increased.
We therefore studied intrinsic properties  of the chain at higher contrasts.

A neural chain with the same parameters as in the previous section was stimulated by a large  stimulus: a ``Gabor patch'' described by (\ref{static_gabor_patch}) with $n_0=20$. The input current $j_0$ varied in the range from 2$\times 10^{-4}$  to 2$\times 10^{-1}$.
The results are plotted in \fig~\ref{fig:gabor-chain-contrast} for six magnitudes of stimulus contrast. To help intuition, we map the magnitudes of $j_0$ on the effective contrasts in the range of  0.001 to 1.00 (see caption of \fig~\ref{fig:gabor-chain-contrast}).

The plots  in \fig~\ref{fig:gabor-chain-contrast} indicate that the effect of stimulus contrast on the intrinsic frequency of the chain depends on stimulus contrast.
At low effective contrasts,  up to 0.02 in these simulations,  the maximum response $r_E(1/n_1)$ of the chain  increases with contrast, but the intrinsic frequency (normalized to 1.0 in the figure) does not change.
At high effective contrasts, above 0.02, increasing the effective contrast leads to an increased maximum response, as before, but is also leads to a larger intrinsic frequency of the network.
For the maximal effective contrast of 1.0 (\ie for the maximum input of $j_0=2\times 10^{-1}$) the maximum of activation is found at the stimulus spatial frequency of 1.43, \ie 43\% higher than at low effective contrasts.

From the analogy of our system with systems of mechanical oscillators, it is expected that the shift of ``resonance frequency'' of the system
could be either positive or negative. 
(One illustration of this analogy is developed in the analysis of anharmonic oscillations in \S28 of \cite{Landau1969mechanics}. 
The numerical simulations summarized in \fig~\ref{fig:gabor-chain-contrast} have so far revealed only an increase of  intrinsic frequency.
Further studies will tell under what conditions the intrinsic frequency of the network decreases with the increase of stimulus contrast.

\section*{Spatiotemporal interference}
\addcontentsline{toc}{section}{Spatiotemporal interference}
\label{sec:spatiotemp}

\noindent 
Now we consider the spatiotemporal interference that arises in the network when the stimuli appear at different locations and at different temporal instants. 
The time course of chain response $r_E(l)$ after extinguishing the stimulus has the form of either a purely exponential temporal decay or a damped temporal oscillation. Here we focus on the  response regime of temporal oscillation, which is similar to the response regime of biological vision, \eg \cite{manahilov1995spatiotemporal,manahilov1998triphasic}.

\subsection*{Response to a short-lived stimulus}
\addcontentsline{toc}{subsection}{Response to a short-lived stimulus}
\label{sec:short_lived_simulations}

\begin{figure}[t!]
	\begin{center}
		\includegraphics[width=.95\textwidth]{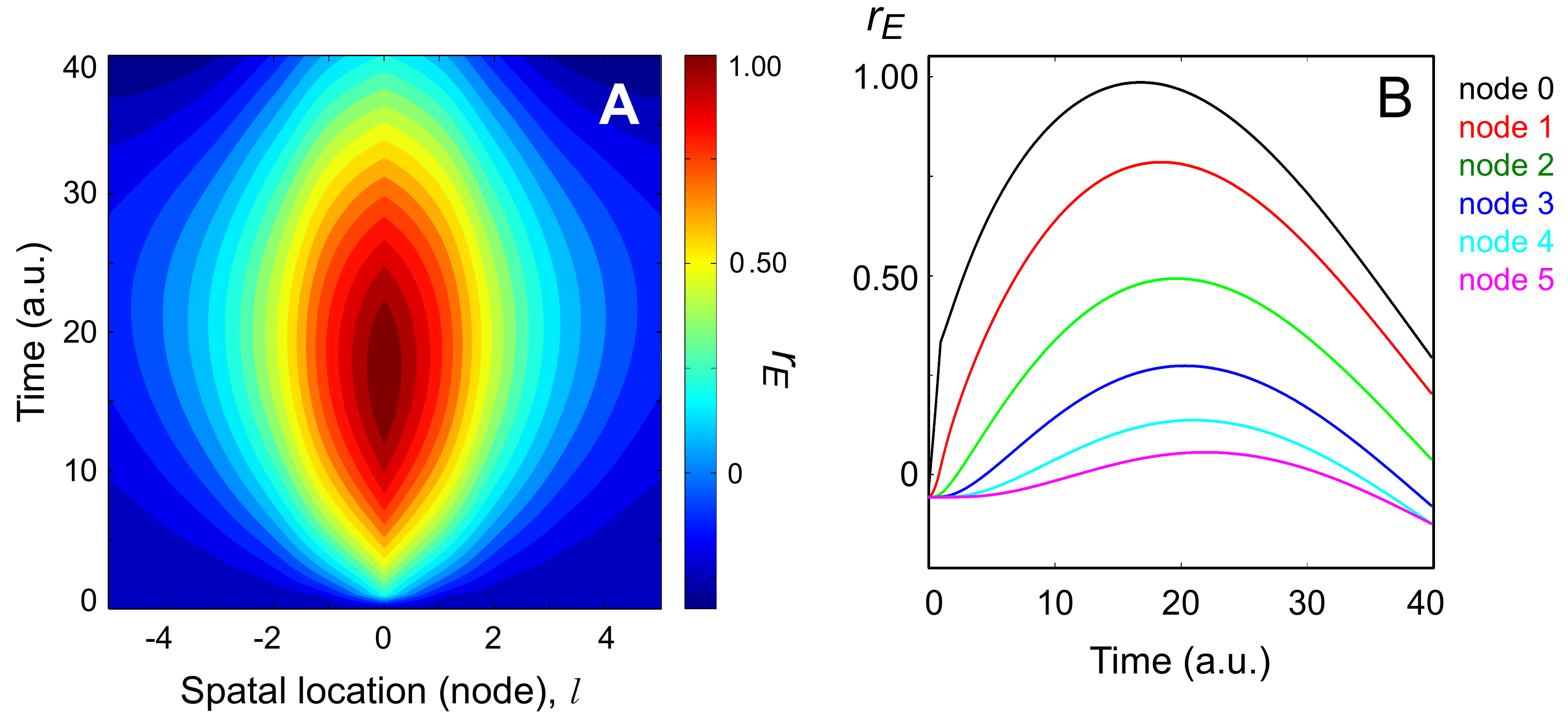}
	\end{center}
	\caption[Spatiotemporal interference (in phase)]{Spatiotemporal interference in-phase. 
		(\textbf{A}) A map of normalized network response $r_E(t,l)$ to the short-lived point stimulus defined in (\ref{eq:short_lived_point}). 
		The abscissa is the spatial coordinate $l$ and the ordinate is the time $t$ after stimulus onset, where $r_E(0,i)=r_I(0,i)=0$. 
		(\textbf{B})~Results of panel~A are shown for the individual nodes $l$ as  functions $r_E(t,l)$, which is $r_E(t,0)$ for the zeroth node, $r_E(t,1)$ for the node $l=1$, etc. The nodes oscillate in phase when~${\cal R}>0$, $t_{\rm stim}=1$.
	}
	\label{fig:spatiotemporal:new}
	\begin{center}
		\includegraphics[width=.95\textwidth]{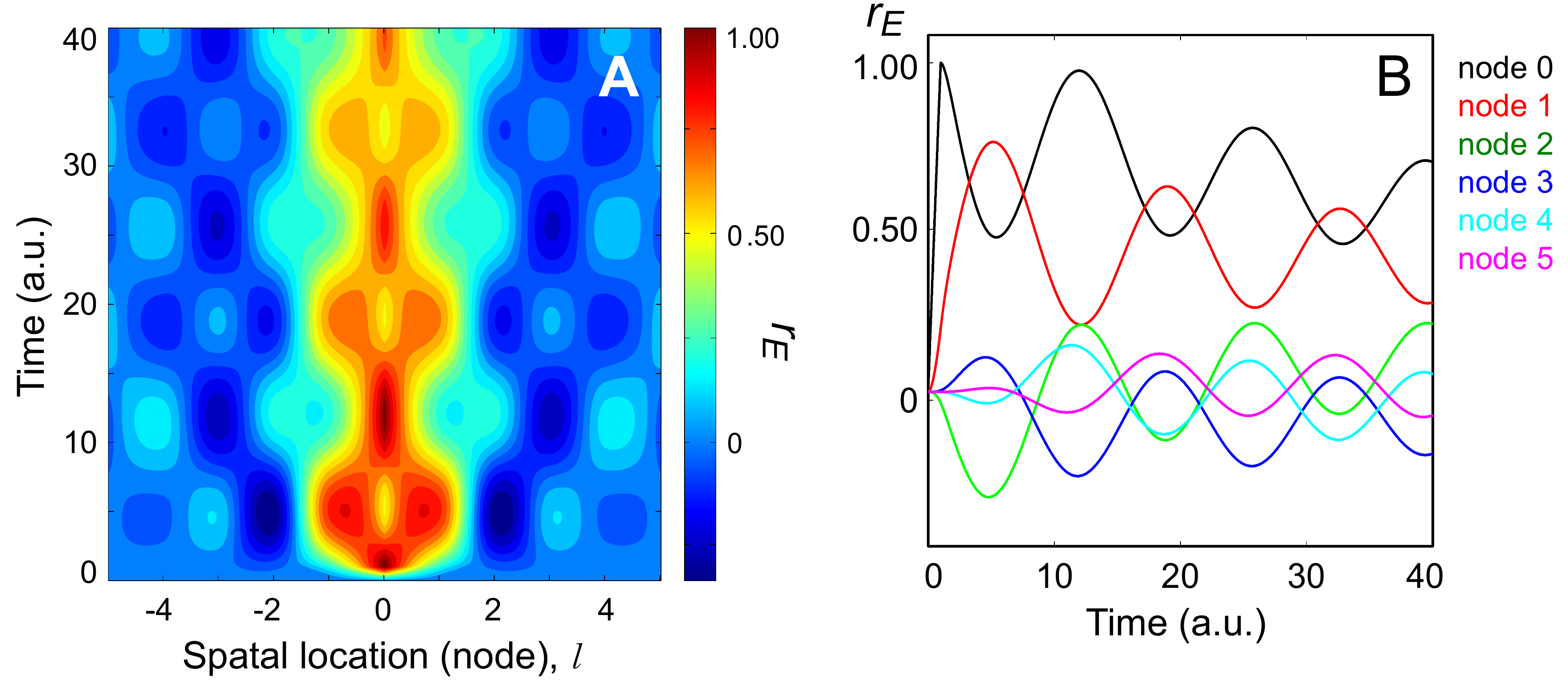}
	\end{center}
	\caption[Spatiotemporal interference (out of phase)]{Spatiotemporal interference out-of-phase.  
		(\textbf{A}-\textbf{B}) The plots as in \fig~\ref{fig:spatiotemporal:new}  for the same stimulus but different chain parameters. 
		The nodes oscillate out of phase when ${\cal R}<0$, $t_{\rm stim}=1$.  (Network parameters for this figure and \fig~\ref{fig:spatiotemporal:new} are listed in the Appendix.) 
	}
	\label{fig:spatiotemporal:new2}
\end{figure}
\begin{figure}[t!]
	\begin{center} 
		\includegraphics[width=.9\textwidth]{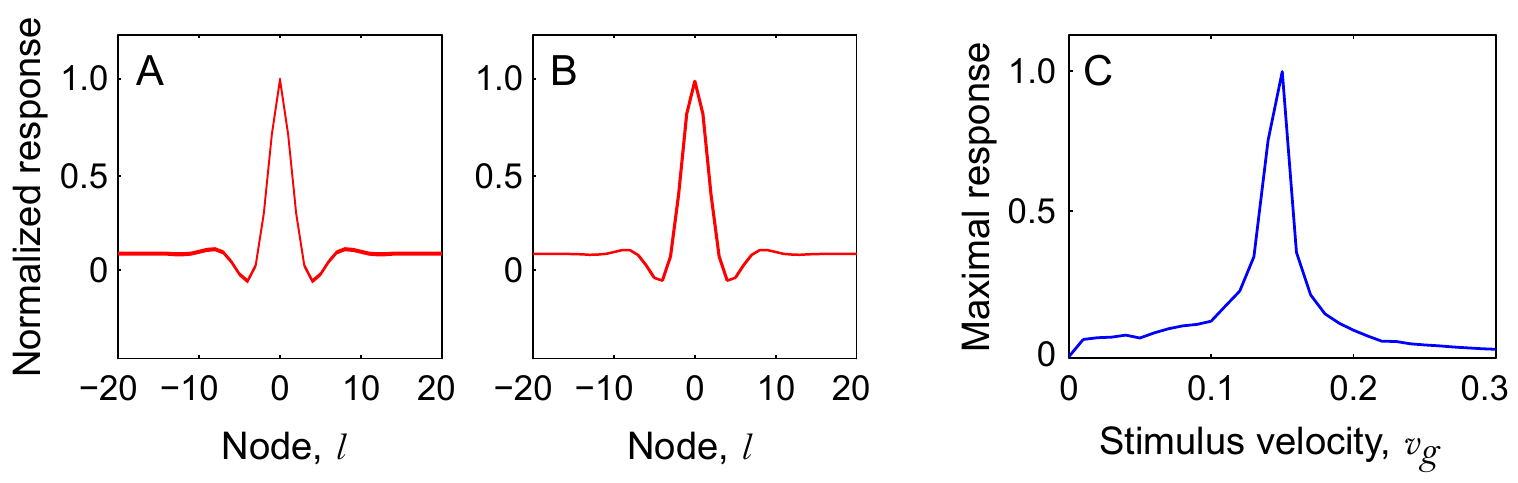}
	\end{center}
	\vspace{-2mm}
	\caption[Steady state responses]{
		({\bf A}--{\bf B})~Normalized steady-state responses for a static ``point'' stimulus using the same network parameters as in \fig~\ref{fig:spatiotemporal:new} for panel~A, and as in \fig~\ref{fig:spatiotemporal:new2} for panel~B.  
		As shown in \figs~\ref{fig:spatiotemporal:new}--\ref{fig:spatiotemporal:new2}, the dynamics of response to short-lived stimuli in these two networks  are significantly different from one another. 
		This is because the response to short-lived stimuli is controlled by parameter $\cal R$ (which has different signs in  \fig~\ref{fig:spatiotemporal:new} and~\ref{fig:spatiotemporal:new2}),
		whereas the steady-state response is controlled by parameters ${\cal T, M, K}$ (which remain the same).  
		%
		({\bf C})~Response to moving stimulus. Normalized maximal response $\max_t[r_E(t, l=0)]$ over the entire observation period on the central node for a drifting Gabor patch (\ref{moving-gabor}) with parameters $n_1=2,\ n_0=20$. 
	} 
	\label{fig:steady}
\end{figure}

\noindent 
We simulated a short-lived point 
excitation of the network with stimulus duration $t_{\rm stim}$, $i_E(t,l)=i_I(t,l)=0$ and $j_0=1$ 
for all $t$ and $l$, except for $l=0, 0<t<t_{\rm stim}$, where
\begin{equation}
i_E(0<t<t_{\rm stim},l=0)=\alpha j_0, i_I(0<t<t_{\rm stim},l=0)=(1-\alpha)j_0.
\label{eq:short_lived_point}
\end{equation}
As we show below, results of these simulations broadly agree with the analysis of network stability presented in Section~{\it Stability of neural chain} (p.~\pageref{sec:stability}).\footnote{
	The chain parameters common to \figs~\ref{fig:spatiotemporal:new}-\ref{fig:steady} were $i_E(t,l)=i_I(t,l)=0$ for all $t$ and $l \neq 0$. 
	For $l=0$,  the system is excited only within the temporal interval $0<t<t_{\rm stim}$: $i_E(0<t<t_{\rm stim},0)=\alpha j,\ i_I(0<t<t_{\rm stim},0)=(1-\alpha)j$. 
	Outside of this interval: $i_E(t>t_{\rm stim},0)=0,\ i_I(t>t_{\rm stim},0)=0$, and $w_{EE}=2,\ w_{IE}=1.5$, ${\cal M}=0.01$, ${\cal Q}=-0.01$, ${\cal T}=-0.8$, ${\cal K}=-0.1$, $\alpha=0.8$. 
	For \fig~\ref{fig:spatiotemporal:new},  ${\cal R}=1$, ${\tilde w}_{EE}=1.3$, and ${\tilde w}_{IE}=1.7$. 
	For \fig~\ref{fig:spatiotemporal:new2},  ${\cal R}=-1$, ${\tilde w}_{EE}=1.5$, and ${\tilde w}_{IE}=1.6$. 
	Simulation time was~40, and the time step was $dt=10^{-6}$. 
	Stimulus duration was unity. 
	The chain was 200 nodes~long. }

For ${\cal R}>0$, the slowest decaying response under $t>t_{\rm stim}$ is found where all the nodes respond in phase (\fig~\ref{fig:spatiotemporal:new}) and the wave of neural excitation propagates with extremely high velocity. 
Notably, the maximum excitation at $r_E(t,l=0)$ is reached long after extinguishing the stimulus. The stimulus was turned off at $t=1$ whereas $r_E$ reached its maximum at $t=20$. 
Such response delays must be taken into account in the studies that use rapidly alternating or short-lived stimuli, \eg  as in \cite{tadin2003perceptual}. 

For ${\cal R}<0$, the longest decay is found in the mode of fast spatial oscillations, \ie for $k=\pi$, resulting in out-of-phase oscillations on the neighboring nodes (\fig~\ref{fig:spatiotemporal:new2}). The out-of-phase oscillations make the spatiotemporal profile of the neural wave more complicated than in the case of ${\cal R}>0$, forming multiple regions of facilitation and suppression in the \{$l, t$\} map in \fig~\ref{fig:spatiotemporal:new2}A even for the single short pulse of stimulation. We therefore expect that different stimuli will generate non-trivial interference patterns. 

To summarize, in spite of the very different spatiotemporal responses of the system to short-lived stimuli in \figs~\ref{fig:spatiotemporal:new} and~\ref{fig:spatiotemporal:new2}, respectively for ${\cal R}>0$ and ${\cal R}<0$, the responses to static stimuli in these two cases are similar to one another. This is because the steady-state response of the network is fully determined by coefficients ${\cal K, T}$ and ${\cal M}$, while ${\cal R}$ has no effect on the linear steady-state response.

The neural waves generated by interference patterns (as in \fig~\ref{fig:spatiotemporal:new2}) have a well-defined velocity, in agreement with the mounting evidence that neural activity propagates through cortical networks in the form of  traveling waves, \eg \cite{bringuier1999horizontal,nauhaus2009stimulus,ray2011network,muller2012propagating,mohajerani2013spontaneous,muller2014stimulus}. 
For example, consider network response to a distributed dynamic stimulus, a drifting Gabor patch:
\begin{equation}
j(l)=j_0\cos(2\pi (l-v_gt/n_1))\exp(-l^2/n_{0}^{2}).
\label{moving-gabor}
\end{equation}
The well-defined velocity of the wave for ${\cal R}<0$ suggests that the interference pattern will generate clear maxima of $r_E(l)$ when the stimulus moves with the same velocity as the intrinsic (``natural'') velocity of the neural wave.  
This result agrees with the numerical stimulations illustrated in \fig~\ref{fig:steady}C. It suggests a simple neural mechanism for tuning the system to stimulus velocity as a special case of spatiotemporal neural-wave interference. This property of neural wave interference makes unnecessary the assumption of  specialized neural circuits for sensing velocity \cite{Reichardt1969movement,vanSanten1985elaborated,watson1985model}. 

In particular, notice that the interference pattern in \fig~\ref{fig:spatiotemporal:new2} has the temporal period of 13, evident in the oscillations of $r_E(t,l)$ shown separately for individual nodes in \fig~\ref{fig:spatiotemporal:new2}B. 
We may therefore estimate the  rate of temporal oscillations $\lambda_+(k)$ of (\ref{lampm}) as $2\pi/13$. 
We also know the characteristic wavelength of spatial oscillations (which is  two nodes since the nodes oscillates out of phase). 
From the identity  $\lambda_s = 2=2\pi/k_s$, where $k_s$ is intrinsic spatial frequency (\ref{eq:lambda_s}) we infer that $k_s =\pi$.
The velocity of neural wave is the ratio of spatial to temporal frequencies of oscillations, which is  $(2\pi/13)/\pi=2/13=0.15$. 
This is the magnitude of the characteristic velocity for the largest value
of node activity observed in our numerical simulations  (\fig~\ref{fig:steady}C). 
This way, the intrinsic velocity of the network estimated from the map in \fig~\ref{fig:spatiotemporal:new2} provides a close approximation to the result of numerical simulations in~\fig~\ref{fig:steady}C.\footnote{Note that, for estimating the velocity, the spatial period should also be taken from the dynamical regime (where $\lambda=2$) and not from the spatial period of the stationary response with static stimuli (as in Fig. 7B, where $\lambda \approx 14$).}

\subsection*{Response to translating motion} 
\addcontentsline{toc}{subsection}{Response to translating motion}
\label{sec:translation}

\noindent
We also studied how the network responds to a moving stimulus. Since a finite time is needed for the neural wave to propagate between network nodes, it is expected that effects of the moving stimulus would be registered in different parts of the system with different delays, akin to  Li\'enard-Wiechert potential in the classical electromagnetic theory \cite{landau2013classical}.

\begin{figure}[t!]
	\begin{center} 
		\includegraphics[width=.55\textwidth]{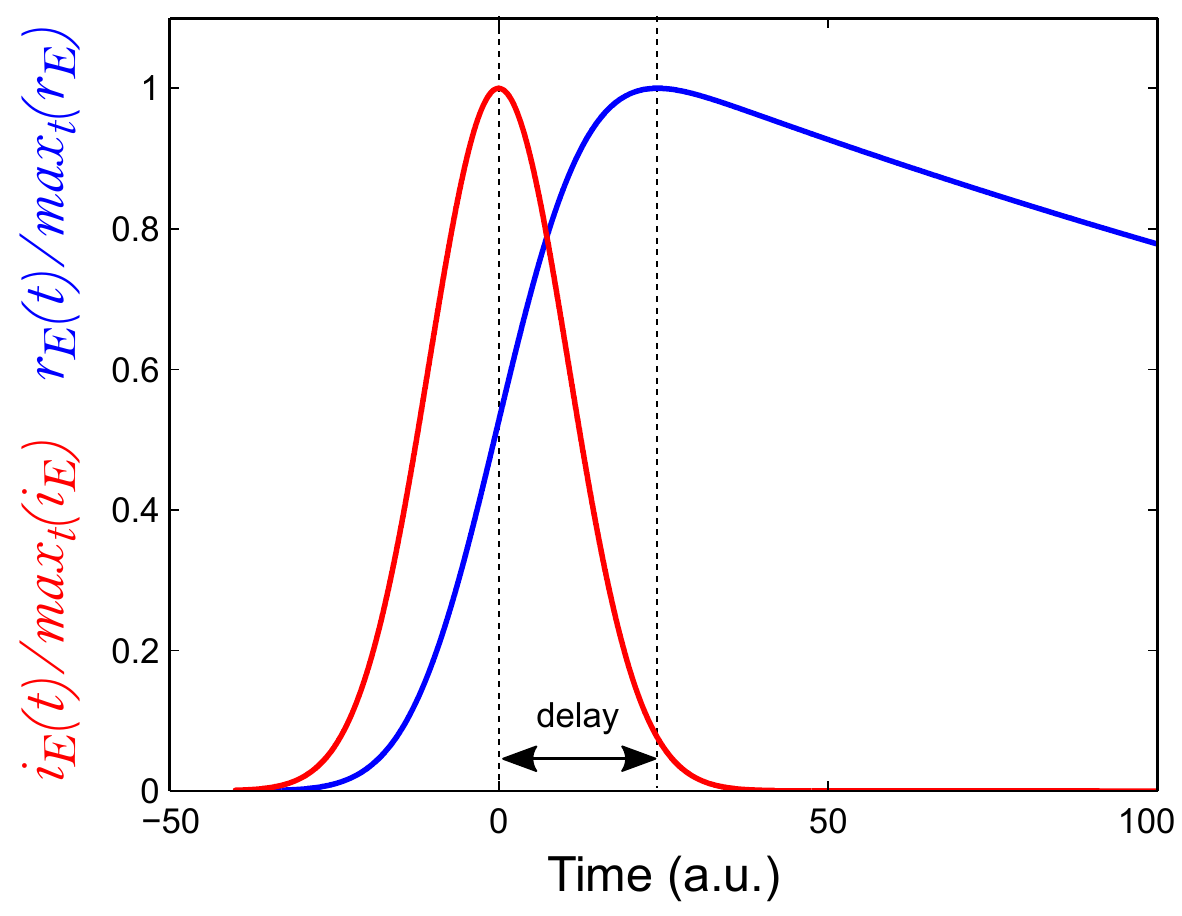} 
	\end{center}
	\vspace{-3mm}
	\caption[Response delay]{
		Time course of the stimulus-driven input $i_E(t,l)$ in red and response $r_E(t,l)$ in blue at the node  $l=0$ of the neural chain 
		with $i_E(t)=i_E(t,i=0)$ and $r_E(t)=r_E(t,l=0)$.
		The ordinate is the magnitude of input current (red) and the response of neural chain (blue), each normalized by its  maximum. 
	} 
	\label{fig:delay}
\end{figure}

For example, consider effects of a small dynamic visual stimulus: a Gaussian spot of light moving with speed $v_g$, thus producing the inputs:
\begin{equation}
j(t,l)=j_0\exp(-(l-v_gt)^2/n_0^2).
\end{equation}
The red curve in \fig~\ref{fig:delay} represents the input to the node  $l=0$. The blue curve represents the network response measured on the same node. 
We defined time $t=0$ to be the instant when the maximum of input occurs at the middle position in the chain, at $l=0$. 
When would the system register the stimulus after the latter has passes the position $l=0$?

Suppose the system registers (``detects'') the stimulus when the firing rate $r_E(t)$ at the position $l=0$ reaches its maximum. 
\fig~\ref{fig:delay} makes it clear that the firing rate keeps growing even after the  stimulus spot has passed the zeroth location. 
The firing rate reaches  its maximum when the input has nearly dropped to zero.  
The magnitude of response delay illustrated in \fig~\ref{fig:delay} depends on the 
control parameters of the network (and thus on the interaction constants)
suggesting that this model can help understanding the delays measured on the neurons, using physiological methods \cite{raiguel1989response,schmolesky1998signal,krekelberg2001neuronal} and measured by psychophysical methods by registering detection thresholds of moving stimuli, as in \cite{nijhawan2002neural,hubbard2005representational}.\footnote{
	The parameters of simulations for \fig~\ref{fig:delay} were  ${\tilde w}_EE= 1.5, {\tilde w}_{IE}= 1.6, {\tilde w}_{II}= 1.579, {\tilde w}_{EI}= 1.496, w_{IE}= 1.5, w_{EE}= 2, w_{II}= 0.901, w_{EI}= 1.317, \tau_E= 1.583, \alpha=0.8, v_g=0.2, n_0=3$. 
}

\section*{Spatial dynamics in a two-dimensional neural array} 
\addcontentsline{toc}{section}{Spatial dynamics in a two-dimensional neural array}
\label{sec:2d}

\subsection*{Model of a two-dimensional neural array}
\addcontentsline{toc}{subsection}{Model of a two-dimensional neural array}

Above, we have investigated a model of one-dimensional (1D) neural chain with nearest-neighbor coupling. 
Such a model can describe interactions of stimuli shaped as parallel stripes or lines. 
To its advantage, the 1D~model can be  studied analytically, at least in part,  greatly simplifying the task of finding the parameters that describe qualitatively different regimes of  network function. 
Yet, a general description of such systems  should  be able to also predict the patterns generated by two-dimensional (2D) stimuli, for which we need to generalize the model to 2D arrangements of the nodes, to which we turn presently.

Neural arrays of two spatial dimensions allow for multiple network geometries and manners of  node coupling.  
Here we present first steps in the analysis of such 2D neural array, aiming to demonstrate how our approach of neural wave interference applies to such systems and how it  helps interpreting  results of experiments in visual psychophysics.

The present model is a square lattice of nodes each containing one excitatory neuron and one inhibitory neuron connected as in \fig~\ref{fig:ISN}A. 
As before, we only consider nearest-neighbor 
coupling, which in the 2D array we implement along the sides and diagonals of each cell of the lattice. 
The relative strengths of coupling along the sides and diagonals is controlled by parameter $\beta$, which is assumed to be the same for all the nodes. 

This model can be written as 
{\small
	\begin{eqnarray}
	\tau_E\frac{dr_E(l,m)}{dt}&=&-r_E(l,m)\nonumber \\ &+& g\left(w_{EE}r_E(l,m)+{\tilde w}_{EE}\sum_{\ \ \ \ \ \ E}r_E-w_{EI}r_I(l,m)-{\tilde w}_{EI}\sum_{\ \ \ \ \ \ I}r_I+i_E(l,m,t)\right), \nonumber \\
	\frac{dr_I(l,m)}{dt}&=&-r_I(l,m)\nonumber \\ &+&g\left(w_{IE}r_E(l,m)+{\tilde w}_{IE}\sum_{\ \ \ \ \ \ E} r_E-w_{II}r_I(l,m)-{\tilde w}_{II}\sum_{\ \ \ \ \ \ I} r_I+i_I(l,m,t)\right), \nonumber \\
	\label{main-plane}
	\end{eqnarray}
}
where 

\vspace{.1in}
\noindent
$\sum_{E}r_E=r_E(l+1,m)+r_E(l-1,m)+r_E(l,m+1)+r_E(l,m-1)+\beta [r_E(l+1,m+1)+r_E(l+1,m-1)+r_E(l-1,m+1)+r_E(l-1,m-1)]$ 

\vspace{.1in}
\noindent
and 

\vspace{.1in}
\noindent
$\sum_{I}r_I=r_I(l+1,m)+r_I(l-1,m)+r_I(l,m+1)+r_I(l,m-1)+\beta[r_I(l+1,m+1)+r_I(l+1,m-1)+r_I(l-1,m+1)+r_I(l-1,m-1)]$. 

\vspace{.1in}
In the following, we consider two cases of  neural fields generated by optical stimuli in this model. 
First is the neural field generated by a static point stimulus.
Such an ``elementary" neural field  can be thought of as a building block of the responses to complex stimulus configurations in the linear regime of the network, as in the studies of the threshold of visibility at low luminance contrasts. 
Second is the neural field generated by a static 2D spatial arrangement of stimuli forming a simple shape,
which produces a nontrivial response.  
The stability conditions for the neural array are analyzed in the Appendix. 

\subsection*{Two-dimensional neural field of a point stimulus}
\addcontentsline{toc}{subsection}{Two-dimensional neural field of a point stimulus}

Following the stability analysis (described in the Appendix), we selected coefficients 
$w_s$, ${\tilde w}_s$ and $\beta$ 
from the region of network stability. 
As the ``point stimulus,'' we considered the activation $j(l=0,m=0)=j_0$, while $j(l,m)=0$ was set for all other locations $l$, $m$.  
As before, we defined $i_{E}(l,m)=\alpha j(l,m)$ and $i_I(l,m)=(1-\alpha)j(l,m)$. 
From the analytical considerations described in the Appendix,  
we know that the solution will have the form of either damped oscillation (for ${\cal M/K}<0$ and $-2-2\beta<{\cal T}<2-2\beta$) 
or exponential decay similar to that described in \fig~\ref{fig:spatial_greens_functions}C.
We concentrate on the more relevant case of damped oscillation.

The network is expected to yield a significant response when the period of neural wave oscillations is considerably longer than the distance between network nodes. 
Using the results of stability analysis described in the Appendix, we set ${\cal T}$ close to $-2-2\beta=-2.8$ (\ie $\beta=0.4$), and we limited ${\cal M}$ to small magnitudes in order to avoid large damping (in full analogy to Section~{\it Static neural waves generated by point stimulus}, p.~\pageref{sec:static_point}). 

\begin{figure}[h!]
	\centering
	\includegraphics[width=0.9\textwidth]{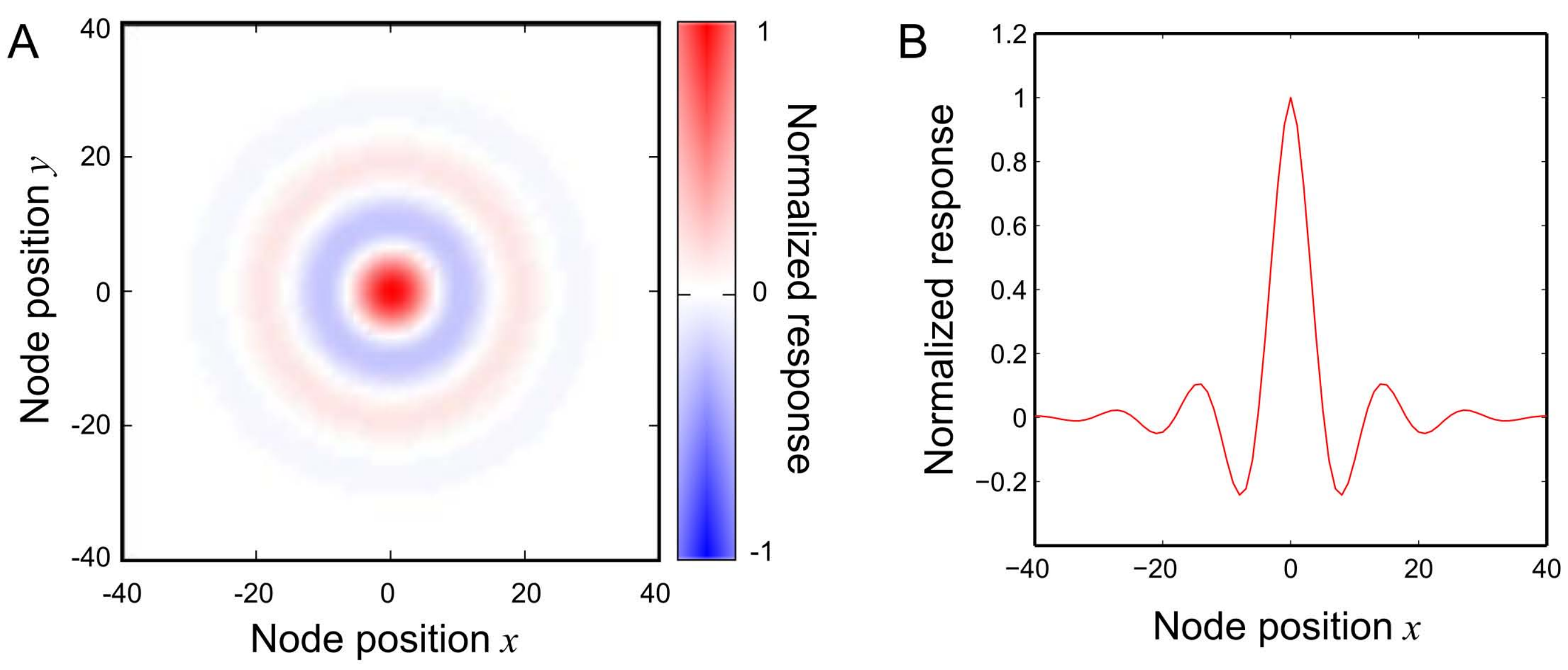}
	\vspace{1mm}
	\caption[Response of a  two-dimensional  neural array to a point stimulus]
	{\small	
		Linear response $r_E(l,m)$ of a two-dimensional neural array. 
		({\bf A})~A two-dimensional map of the normalized response $r_E/\max{r_E}$ to a point stimulus
		applied at the node position~$(0,0)$.  
		({\bf B})~A horizontal section of the response map at the node~0.
		The simulations were performed on the array of 201$\times$201 nodes. 
		The map in~A shows the response subset on 121~nodes, centered on (0, 0),
		and the slice in~B shows the response of a smaller set on 81~nodes, to improve the visibility of detail.
	} 
	\label{fig:2D_point} 
\end{figure}

One result of the neural field generated by a  point stimulus is displayed in \fig~\ref{fig:2D_point}. 
The neural field in \fig~\ref{fig:2D_point}A has the spatial period of about 14~nodes, yielding several salient ``rings" of excitation and inhibition.  
Concentric regions of the strongest inhibition and the strongest facilitation are found at the distances of, respectively, $\Delta {\cal L}=7$ and $2\Delta {\cal L}=14$ from the excited point. 

Just as we indicated in Section~{\it Static neural waves generated by point stimulus} (p.~\pageref{sec:static_point}), the alternating regions of excitation and inhibition formed by the point stimulus resemble the spatial oscillations described in psychophysical studies \cite{manahilov1995spatiotemporal,manahilov1998triphasic} 
using one-dimensional variations of luminance, here predicted to generalize to two spatial dimensions.   
As in the one-dimensional model, system response to complex two-dimensional stimuli at low stimulus contrasts (of the sort used in the next section) should be predicted from superposition of such ``elementary'' neural fields.
In other words, the neural field shown in \fig~\ref{fig:2D_point}A can be used as a convolution kernel. 

\subsection*{Neural field of an elliptic ring stimulus}
\addcontentsline{toc}{subsection}{Neural field of an elliptic ring stimulus}

\begin{figure}[t!]	
	\centering
	\includegraphics[width=.8\textwidth]{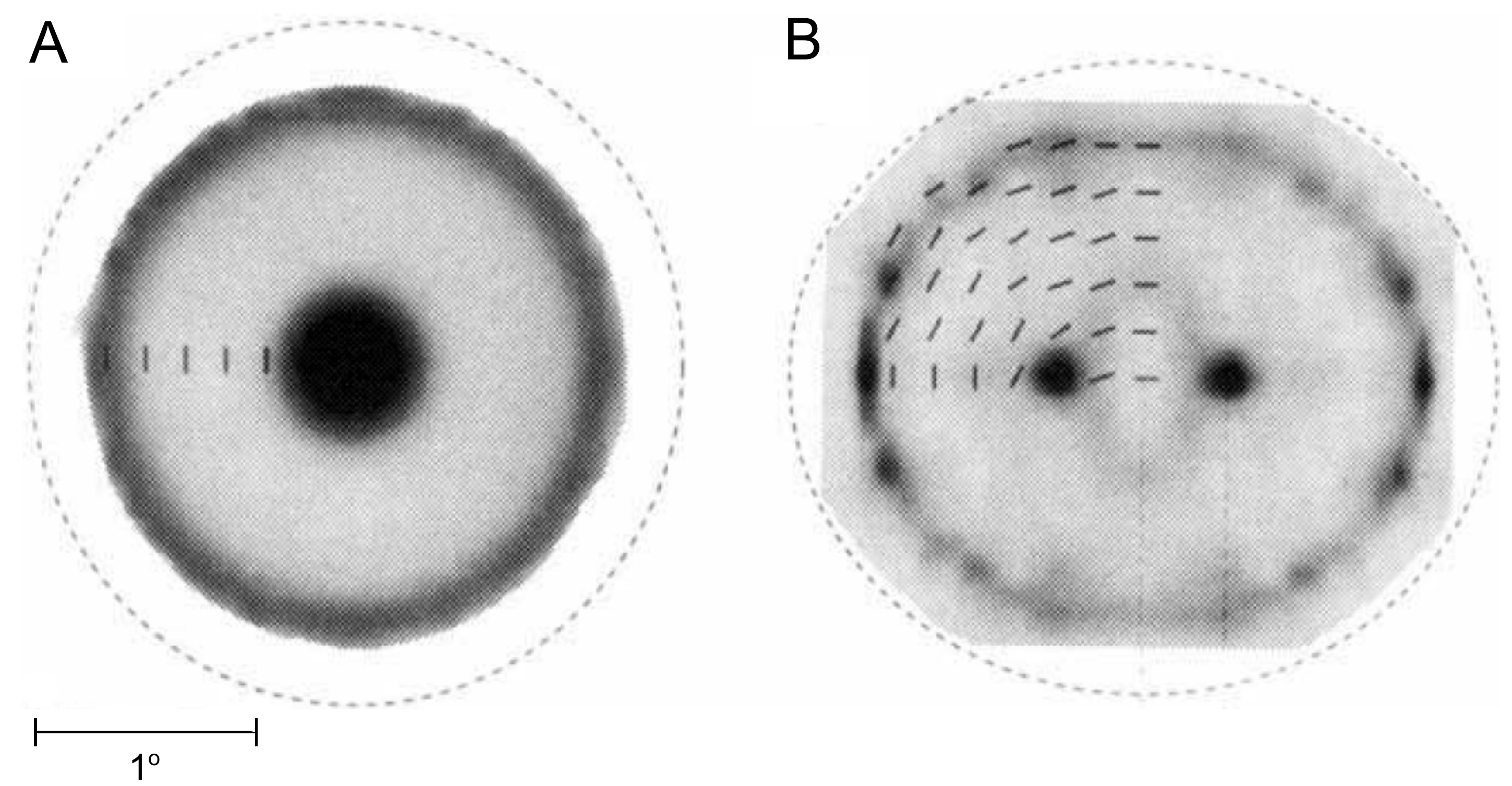} 
   \vspace{2mm}
	\caption[Map of measured changes of senstivity induced by ring stimuli]
	{\small  
		Maps of sensitivity change induced by circular (A) and elliptical (B)  high-contrast ring stimuli \cite{kovacs1994perceptual}.  The dashed contours represent the locations of stimuli: strings of ``Gabor patches" aligned with the contour. The gray-level images represent the change of contrast sensitivity induced by the contour, measured at the locations and orientations of low-contrast probes indicated by the short dark lines. The darker gray levels in the maps represent the larger increments of sensitivity, with the largest increment of 0.4~log units. \emph{Adapted from \cite{kovacs1994perceptual}.}
	}
	\label{fig:kovacs}
\end{figure}

Interference of neural fields generated by multiple elementary stimuli 
may form a peculiar spatial pattern of excitation and inhibition.  
The pattern will consist of regions where the potential stimuli can be facilitated or suppressed by the inducing stimulus. 
The facilitation and suppression can be revealed by measuring the contrast threshold of a low contrast probing stimulus placed at multiple locations in the region of interest~(as in \fig~\ref{fig:kovacs}). 
Indeed, if such a probe appears in the facilitated or suppressed regions, its contrast threshold should be, respectively, lower or higher than in the absence of the inducer. 
This way, the neural field of the complex two-dimensional stimulus can be experimentally observed and compared with  predictions of the model.

We studied the neural field generated by a stimulus that has the shape of ellipse, which we modeled  using the network input  $j(l,m)=0$ for any $l,m$ except of $l,m$ satisfying the inequality:
\[ |\sqrt{l^2/R_{1}^{2}+m^2/R_{2}^{2}}-1|<\Delta R/\sqrt{R_1R_2}, \]
where $R_1$ and $R_2$ are the radii, $\Delta R$ is the thickness of the elliptic ring (required for generating a ring on a discrete array of nodes), and $j(l,m)=j_0$. 

The four versions of the inputs $j(l,m)$ used in the simulations are shown on the right side  of \fig~\ref{fig:2D_elipse}.  
The radii $R_1$ and $R_2$ selected for the simulations were multiples of the half period $\Delta{\cal L}=7$ of the 
intrinsic neural-field oscillations discovered using a point stimulus (described in the previous section; \fig~\ref{fig:2D_point}), 
in order to obtain salient effects of constructive and destructive interference. 
The  neural fields generated by these stimuli were obtained by numerical simulations of (\ref{main-plane}). 
Four examples of the results, along with the corresponding stimuli,  are shown at left in \fig~\ref{fig:2D_elipse}: 

\begin{figure}[t!]	
	\begin{minipage}[c]{0.60\textwidth}
		\centering
		\includegraphics[width=\textwidth]{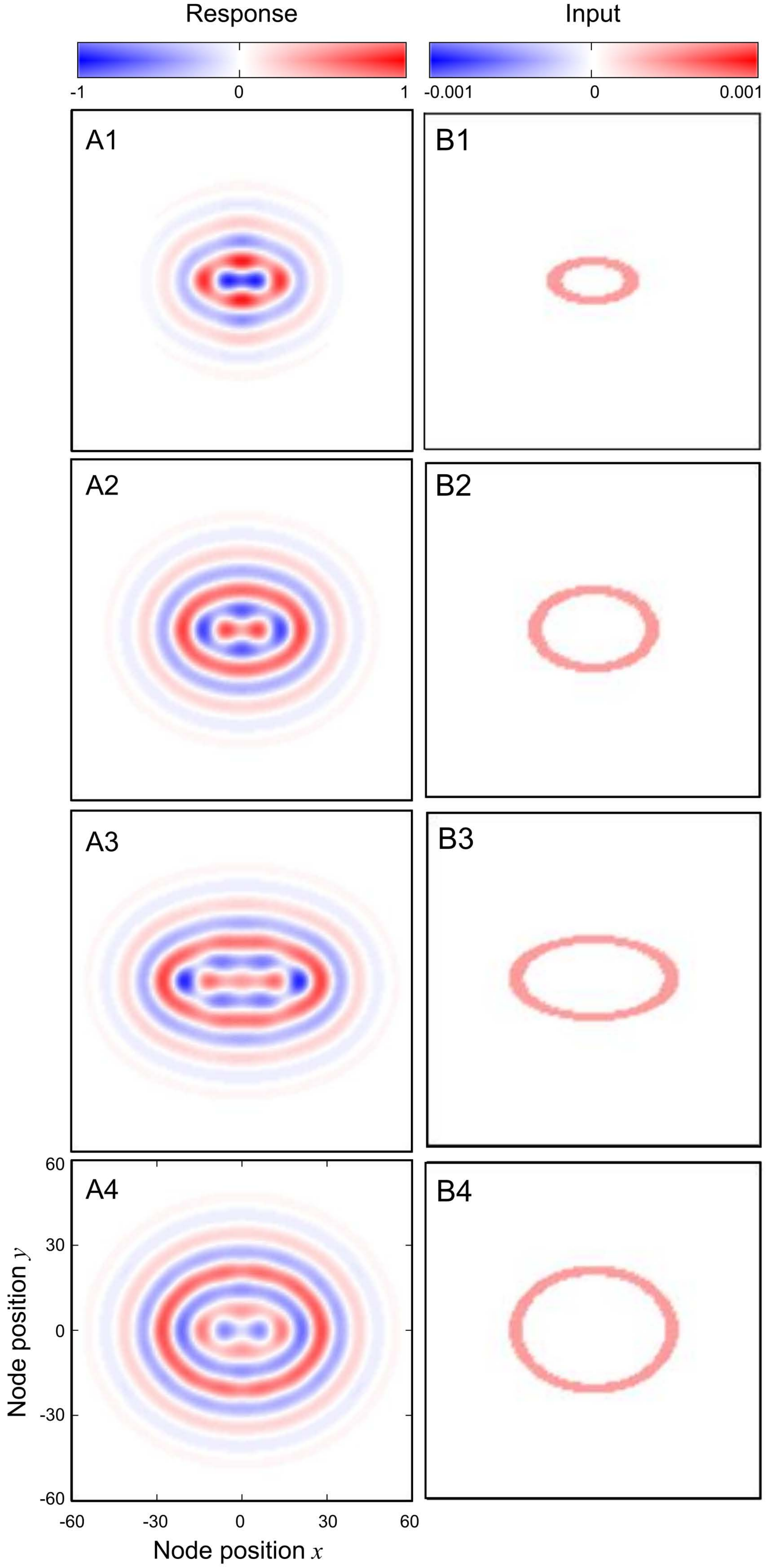} 
	\end{minipage}
	\hfill
	\begin{minipage}[c]{0.35\textwidth}
		\caption[Response of a two-dimensional neural array to elliptic ring stimuli]
		{   \small 
			Response of a two-dimensional neural array to elliptic ring stimuli. 
			The maps at left (panels~A) depict the responses $r_E(l,m)$ of a two-dimensional neural array 
			(with the same parameters as in \fig~\ref{fig:2D_point}) to the elliptic ring stimuli 
			shown within the same rows at right ((``inputs," panels~B). 
			The color maps for the responses and inputs are defined separately in the color bars above the panels at left and right. 
			The four stimuli had the following radii ($R_1,R_2$), top to bottom: (14,7), (21,14), (28, 14), (28,21). 	
			All the panels are drawn in the same coordinates of node position ($x, y$) as in the panel~A4.  
		} 
		\label{fig:2D_elipse}
	\end{minipage}
\end{figure}

\begin{description}
	\item 
	For a small highly elongated elliptic stimulus (panel~B1), mostly destructive interference is found inside the ellipse (A1), where the neural activity is suppressed.
	
	\item 
	When increasing the stimulus semi-axes: $R_1=3{\cal L}, R_2=2{\cal L}$ (B2), the competition between excitation and inhibition produces excitation regions in the two foci of the ellipse (A2). 
	
	\item 
	Increasing one of the stimulus semi-axes yet further: $R_1=4{\cal L}, R_2=2{\cal L}$ (B3) creates three localized regions of facilitation inside the ellipse, along its major axis (A3).
	
	\item 
	And increasing the  other semi-axis: $R_1=4{\cal L}$, $R_2=3{\cal L}$ (B4) creates a zone of excitation inside the stimulus in the shape of an ellipse (A4). 
\end{description}
These results suggest that the simplified architecture of neural connectivity with nearest-neighbor coupling can serve as a useful modeling framework for understanding how two-dimensional stimulus configurations modulate pattern visibility, as in \fig~\ref{fig:kovacs},
and in other studies that engage neural interactions between spatial locations in more than one spatial dimension, \eg  \cite{field1993contour,kovacs1993closed,itti1998model,yen1998extraction,kapadia2000spatial,rosenthal2007spatiotemporal,pelli2009grouping,mcmanus2011adaptive,wagemans2012century}.

\section*{Conclusions}
\addcontentsline{toc}{section}{Conclusions}

\noindent 
We investigated interference of  neural waves in inhibitory-excitatory neural chains with nearest-neighbor coupling.  
We defined conditions of stability in such systems  with respect to corrugation perturbations and derived the control parameters that determine how such systems respond to static, short-lived, and moving stimuli. 
We found that the interference patterns generated in such networks endow the system with several properties commonly observed in biological vision, including  selectivity of the network for spatial and temporal frequencies of stimulus intensity modulation, selectivity for stimulus velocity,  ``lateral'' interactions between spatially separate stimuli, and predictable delays in system response to static and moving stimuli. 
We also investigated interference of  neural waves in two-dimensional generalizations of such networks and found that two-dimensional neural fields form patterns  whose properties resemble the properties of contrast sensitivity patterns induced across spatial location by two-dimensional visual stimulus configurations.  
It is plausible that neural wave interference is responsible for some of the visual phenomena that had been attributed to specialized neural circuits and studied using distinct modeling frameworks, including the frequency tuning of neural circuits, their direction selectivity, contrast normalization, and ``lateral" (center-surround) spatial interactions.


\setcounter{equation}{0}
\renewcommand\theequation{A.\arabic{equation}}

\setcounter{section}{0}
\renewcommand\thesection*{\Alph{section}}

\section*{Appendix: Stability of  2D neural arrays}
\addcontentsline{toc}{section}{Appendix: Stability of  2D neural arrays}

\label{sec:array_stability}

Similar to our analysis of the one-dimensional model, we first study the stability of the solution $r_E=0,\ r_I=0$ for $i_E=i_I=0$ for all $l$ and $n$
in the linearized equations (\ref{main-plane}) with $g({\cal X})={\cal X}$. 
We use perturbations of the same form as in the analysis of the one-dimensional system (Section~{\it Stability of neural chain}, p.~\pageref{sec:stability}):
\begin{eqnarray}
r_E=\Delta_E e^{\lambda t+ik_x l+ik_y m} \nonumber \\
r_I= \Delta_I e^{\lambda t +ik_x l+ik_y m}.
\label{plane-perturb}
\end{eqnarray}
Given the spatial decay rate $\lambda$ and the wave vector components $k_x$, $k_y$, we derive a set of algebraic equations for wave amplitudes $\Delta_E$ and $\Delta_I$:
\begin{eqnarray}
(w_{EE}+2{\tilde w}_{EE} f(k_x,k_y)-1-\tau_E\lambda)\Delta_E-(w_{EI}+2{\tilde w}_{EI} f(k_x,k_y))\Delta_I=0 \nonumber \\
(w_{IE}+2{\tilde w}_{IE} f(k_x,k_y))\Delta_E-(w_{II}+2{\tilde w}_{II} f(k_x,k_y) +1+\lambda)=0.
\label{plane-perturb1}
\end{eqnarray}
Here we introduce function
\[ f(k_x,k_y)=\cos k_x +\cos k_y + \beta [\cos (k_x+k_y)+\cos(k_x-k_y)], \] 
which reaches its maximum value $\max f=2+2\beta$ at $k_x=k_y=0$, and its minimum value: either $\min f=-2+2\beta$ 
at $k_x=k_y=\pi$ for $\beta<1/2$, or 
$\min f=-2\beta$ 
at $k_x=0, k_y=\pi$ for $\beta>1/2$. That is, different stability regions are expected for different magnitudes of parameter $\beta$. 

To simplify, we consider the rapidly decaying coupling ($\beta<1/2$) where the interactions of nearest-neighbor nodes along lattice diagonals is considerably weaker than the interactions along lattice sides (since the diagonal distance is longer by factor $\sqrt{2}$ than distance along the  sides). Following the approach we introduced in the analysis  of the one-dimensional system, we derive two stability conditions. One reduces to
\begin{equation}
w_{EE}-1-\tau_{E}w_{II}-\tau_E+4|{\tilde w}_{EE}-\tau_{E}{\tilde w}_{II}|+4\beta({\tilde w}_{EE}-\tau_{E}{\tilde w}_{II})<0,
\end{equation}
while the other can be written as:
\begin{eqnarray}
{\cal F}={\cal M}
-{\cal K}\left( f(k_x,k_y)+{\cal T}\right)^2>0.
\label{plane-stab-cond1}
\end{eqnarray}

The most ``dangerous'' point in the $k_x, k_y$ space 
is expected where a negative rate $\lambda(k_x,k_y)$ can become positive (\ie where the real part of $\lambda$ crosses zero), resulting in an instability.
This point occurs at the minimum of the function ${\cal F}$, where the function changes its sign.  
Since~${\cal F}$ is parabolic with respect to variable $f$, which changes in the interval $-2+2\beta<f\leq 2+2\beta$, it can attain its minimum either at
$f=\pm 2+2\beta$ or at its stationary point $f={\cal T}$ (where $\partial {\cal F}/\partial f=0$) which is the minimum for ${\cal K}<0$. 

Comparing values of ${\cal F}$ at the three ``dangerous'' points results in the following set of stability conditions, in which include the ``dangerous" $k_x$ and $k_y$:

\noindent
{ \footnotesize
	\begin{eqnarray}
	{\cal M}&-&{\cal K}(-2+2\beta+{\cal T})^2=(w_{II}+1)(1-w_{EE})+w_{EI}w_{IE}-4({\tilde w}_{II}{\tilde w}_{EE}-{\tilde w}_{EI}{\tilde w}_{IE})(2-2\beta)^2
	\nonumber \\&+&2({\tilde w}_{EE}(w_{II}+1)+
	{\tilde w}_{II}(w_{EE}-1)-{\tilde w}_{EI}w_{IE}-{\tilde w}_{IE}w_{EI})(2-2\beta)>0
	\nonumber \\ &{\rm if}&\ \ {\cal K}>0,\ {\cal T}<-2\beta,\ k_x=k_y\approx -\pi, \label{stab-2d-1}
	\end{eqnarray}
	\begin{eqnarray}
	{\cal M}&-&{\cal K}(2+2\beta+{\cal T})^2=(w_{II}+1)(1-w_{EE})+w_{EI}w_{IE}-4({\tilde w}_{II}{\tilde w}_{EE}-{\tilde w}_{EI}{\tilde w}_{IE})(2+2\beta)^2\nonumber \\
	&-&2({\tilde w}_{EE}(w_{II}+1)+
	{\tilde w}_{II}(w_{EE}-1)-{\tilde w}_{EI}w_{IE}-{\tilde w}_{IE}w_{EI})(2+2\beta)>0\nonumber\\
	&{\rm if}&\ \ {\cal K}>0,\ {\cal T}>-2\beta,\ k_x=k_y\approx 0,
	\label{stab-2d-2}
	\end{eqnarray}
	\begin{eqnarray}
	{\cal M}&-&{\cal K}(-2+2\beta+{\cal T})^2=(w_{II}+1)(1-w_{EE})+w_{EI}w_{IE}-4({\tilde w}_{II}{\tilde w}_{EE}-{\tilde w}_{EI}{\tilde w}_{IE})(2-2\beta)^2\nonumber \\
	&+&2({\tilde w}_{EE}(w_{II}+1)+
	{\tilde w}_{II}(w_{EE}-1)-{\tilde w}_{EI}w_{IE}-{\tilde w}_{IE}w_{EI})(2-2\beta)>0\nonumber \\
	&{\rm if}&\ \ {\cal K}<0,\ {\cal T}>2-2\beta,\ k_x=k_y\approx -\pi,
	\label{stab-2d-3}
	\end{eqnarray}
	\begin{eqnarray}
	{\cal M}&-&{\cal K}(2+2\beta+{\cal T})^2=(w_{II}+1)(1-w_{EE})+w_{EI}w_{IE}-4({\tilde w}_{II}{\tilde w}_{EE}-{\tilde w}_{EI}{\tilde w}_{IE})(2+2\beta)^2\nonumber \\
	&-&2({\tilde w}_{EE}(w_{II}+1)+
	{\tilde w}_{II}(w_{EE}-1)-{\tilde w}_{EI}w_{IE}-{\tilde w}_{IE}w_{EI})(2+2\beta)>0\nonumber\\
	&{\rm if}&\ \ {\cal K}<0,\ {\cal T}<-2-2\beta,\ k_x=k_y\approx -0
	\label{stab-2d-4}
	\end{eqnarray}
	\begin{eqnarray}
	{\cal M}&=&(w_{II}+1)(1-w_{EE})+w_{EI}w_{IE}+\frac{({\tilde w}_{EE}(w_{II}+1)+
		{\tilde w}_{II}(w_{EE}-1)-{\tilde w}_{EI}w_{IE}-{\tilde w}_{IE}w_{EI})^2}{4({\tilde w}_{II}{\tilde w}_{EE} -{\tilde w}_{EI}{\tilde w}_{IE})}>0\nonumber\\
	&{\rm if}&\ \ {\cal K}<0,\ -2-2\beta<{\cal T}<2-2\beta,\ f(k_x,k_y)=-{\cal T}.
	\label{plane-second-stability}
	\end{eqnarray}
}

The ``dangerous'' values of $k_x$ and $k_y$ determine the spatial scale of oscillations at which the instability is expected to occur. 
The conditions most relevant for description of extended stimuli are found where the intrinsic spatial wavelength of the network 
is long in comparison to the inter-node distance and where the network response varies smoothly with \emph{stimulus} parameters.
These conditions are met only near the boundary of stability (\ref{plane-second-stability}), thus considerably restricting the region of relevant parameters of the network. 
Outside of this region, the intrinsic wavelength  
$\propto 1/\sqrt{k_x^2+k_y^2}$ does not depend on the coupling coefficients. 
In the latter case, the intrinsic wavelength can be too large: as in (\ref{stab-2d-2}) and~(\ref{stab-2d-4}), or too small: as in (\ref{stab-2d-1}) and~(\ref{stab-2d-3}).

Here we  focus on the stability region defined in (\ref{plane-second-stability}), where the condition that determines the intrinsic wavelength is $f(k_x,k_y)=-{\cal T}$. 
We consider the case of  small wave vectors $(k_x,k_y)$, where the neural wave changes from node to node smoothly, and where the expression for the intrinsic $k_x$ and $k_y$ can be  simplified further~to 
\[k^2=k_x^2+k_y^2=({\cal T}+2+2\beta)/(1/2+\beta).   \; \Box \]

\section*{Acknowledgments}
We thank Thomas D. Albright, Monika Jadi, Lyle Muller, Ambarish Pawar, Joseph Snider, and Gene Stoner for illuminating discussions. The research was supported in part by the Leverhulme Trust (Savel'ev) and National Institutes of Health Grant EY018613 (Gepshtein).


\addcontentsline{toc}{section}{References}

\singlespacing
%


\end{document}